\documentclass[fleqn,usenatbib]{mnras}

\usepackage{newtxtext,newtxmath}

\usepackage[T1]{fontenc}

\DeclareRobustCommand{\VAN}[3]{#2}
\let\VANthebibliography\thebibliography
\def\thebibliography{\DeclareRobustCommand{\VAN}[3]{##3}\VANthebibliography}

\usepackage{graphicx}	
\usepackage{amsmath}	

\usepackage{graphicx}	
\usepackage{amsmath}	
\usepackage{mathtools,orcidlink}

\newcommand{\oh}{$\rm12+\log(O/H)$}
\newcommand{\no}{$\rm\log(N/O)$}
\newcommand{\W}{$\lambda$}
\newcommand{\WW}{$\lambda\lambda$}
\newcommand{\ph}{\phantom{1}}
\newcommand{\kms}{$\rm km~s^{-1}$}
\newcommand{\ccm}{$\rm~cm^{-3}$}

\title[New Ionization Models and Nitrogen Excess]{New Ionization Models and the Shocking Nitrogen Excess at $z>5$}

\author[S. R. Flury et al]{
Sophia R. Flury\orcidlink{0000-0002-0159-2613},$^{1,2}$\thanks{E-mail: sflury@roe.ac.uk}
Karla Z. Arellano-C{\'o}rdova\orcidlink{0000-0002-2644-3518},$^{1}$
Edward C. Moran, $^{3}$
and Alaina Einsig\orcidlink{0009-0002-8732-6777}$^{2,3}$
\\
$^{1}$Institute for Astronomy, University of Edinburgh, Royal Observatory, Edinburgh, EH9 3HJ, UK\\
$^{2}$Department of Astronomy, University of Massachusetts Amherst, Amherst, MA 01002, USA\\
$^{3}$Department of Astronomy, Van Vleck Observatory, Wesleyan University, 96 Foss Hill Drive, Middletown, CT 06459, USA
}

\date{Accepted XXX. Received YYY; in original form ZZZ}

\pubyear{\the\year{}}

\begin{document}
\label{firstpage}
\pagerange{\pageref{firstpage}--\pageref{lastpage}}
\maketitle

\begin{abstract}
The new era of galaxy evolution studies hearkened in by \emph{JWST} has led to the discovery of $z>5$ galaxies exhibiting excess nitrogen with \no$\sim$1 dex or more than expected from \no-\oh\ trends in the local Universe. A variety of novel enrichment pathways have been presented to explain the apparent nitrogen excess, invoking a wide range of processes from very massive stars to stripped binaries to fine-tuned star-formation histories. However, understanding the excitation mechanism responsible for the observed nebular emission is necessary to accurately infer chemical abundances. As of yet, the ionization sources of these galaxies have not been thoroughly explored, with radiative shocks left out of the picture.
We present a suite of homogeneous excitation models for star-forming galaxies, active galactic nuclei, and radiative shocks, with which we explore possible explanations for the apparent nitrogen excess.
We propose new BPT-style diagnostics to classify galaxies at $z>5$, finding that, when combined with \ion{O}{iii}] \WW1660,66 and \ion{He}{ii} \W1640, \ion{N}{iii}] \WW1747-54 / \ion{C}{iii}] \WW1907,09 best selects shock-dominated galaxies while \ion{N}{iv}] \WW1483,86 / \ion{C}{iii}] \WW1907,09 best distinguishes between active black holes and star forming galaxies.
From our diagnostics, we find that slow/intermediate radiative shocks ($v=75$-$150$ \kms) are most consistent with observed UV emission line flux ratios in nitrogen-bright galaxies.
Accounting for the effects of shocks can bring nitrogen estimates into better agreement with abundance patterns observed in the local Universe
and may be attributable to Wolf Rayet populations actively enriching these galaxies with nitrogen and possibly driving winds responsible for these shocks.
\end{abstract}







\section{Introduction} \label{sec:intro}

With the dawn of the \emph{JWST} era, reports of enhanced nitrogen in multiple galaxies at high redshift ($z>5$) have astounded the astronomical community. Notable cases include GHZ2 \citep[$z=12.4$,][]{Castellano2024}, GN-z11 \citep[$z=10.6$,][]{Bunker2023,Senchyna2023}, CEERS-1019 \citep[$z=8.678$][]{Larson2023,MarquesChaves2024}, and GS 3073 \citep[$z=5.55$,][]{Ji2024}. Standard chemical evolution models based on nucleosynthesis and stellar yields struggle to explain high \no\ ratios in general \citep[e.g.,][]{Kobayashi2020}. Explanations of the nitrogen excess at $z>5$ invoke multi-burst star formation scenarios with particularly narrow time windows ($<1$ Myr) during which such high \no\ is attainable \citep{Kobayashi2024}, low-metallicity stripped binary stars in proto globular clusters \citep{Senchyna2023,Isobe2023}, very massive stars \citep[VMS, e.g.,][]{Vink2023,RiveraThorsen2024,Nandal2024a}, or rapidly rotating Population III stars with fine-tuned abundance yields \citep{Nandal2024b}. Moreover, the gas excitation mechanism remains wholly ambiguous. Both GN-z11 and GHZ2 reside ambiguously on the threshold between active galactic nuclei (AGN) and starburst photoionization model predictions \citep{Bunker2023,Castellano2024} with additional evidence suggestive of an AGN \citep{Maiolino2024}. CEERS-1019 exhibits evidence for both VMS \citep[][]{MarquesChaves2024} and a supermassive black hole \citep[SMBH,][]{Larson2023,MarquesChaves2024}.

Clearly, follow-up investigation is necessary to understand the origins of both the nitrogen enhancement and the gas excitation in these esoteric yet fundamentally important galaxies. Among the excitation mechanisms considered in the literature regarding galaxies with reported nitrogen excess, one remains strangely absent: shocks. A variety of phenomena can cause radiative shocks, including supernovae, stellar winds, and outflows \citep[e.g.,][]{Draine1993,Sutherland1993,Dopita1995a,Allen2008}, all of which occur in galaxies across cosmic time and are thus reasonable to consider as an alternative to those proposed in the literature. However, comparing existing published shock and photoionization model libraries comes with several caveats which complicate their simultaneous use to characterize ionization mechanisms, including differences in abundance patterns \citep[e.g.,][]{Nicholls2017} and in atomic data and underlying physics \citep[e.g.,][]{Zhu2023}.

The popular photoionization code {\tt Cloudy} \citep{cloudy17} does not support shock models. As a result, the current libraries of photoionization models in active use, built using {\tt Cloudy}--noteably \citet{Gutkin2016} for star-forming galaxies (SFGs) and \citet{Feltre2016} for AGN--utilize a different implementation of atomic data and radiative transfer and ionization equilibrium solutions than the most contemporary \citet{Alarie2019} shock model library {\tt EOMS}, built using {\tt MAPPINGS V} \citep{Sutherland2017,mappingsV}. The variations between the {\tt Cloudy} and {\tt MAPPINGS V} codes can lead to systematic differences of anywhere from 25 to 300\% in predicted line fluxes \citep[e.g.,][]{Zhu2023}.  None of these libraries use empirical abundance patterns based on O and B stars and are therefore more likely biased by systematics associated with, e.g., scaling with solar abundances or fits to \ion{H}{ii} region measurements \citep[e.g.,][]{Gutkin2016}. Using empirical abundance patterns from young stars can have a significant impact on predicted flux ratios, particularly when attempting to describe the range of observed flux ratios in galaxies \citep{Nicholls2017}. Additionally, most published ionization model libraries do not holistically account for the effects of dust on gas, instead simply offsetting the carbon abundance relative to oxygen to account for depletion onto grains \citep[e.g.,][]{Gutkin2016}. A complete, empirical treatment of dust depletion can affect the gas phase abundance patterns of many elements (notably carbon, silicon, and iron) and the dust content of the gas. The dust and gas content directly affects cooling by line emission from metals, heating by ionization and the photoelectric effect, and the rate at which ionizing energy is absorbed by the gas \citep[e.g.,][]{Dopita2002,Groves2004}. All of these effects have implications for the temperature, density, and ionization structure of model nebulae.

We present a new, uniformly implemented and empirically motivated library of ionization models to explore shocks as an alternative explanation for the observed \ion{C}{iii}], \ion{C}{iv}, \ion{N}{iii}, \ion{N}{iv}], \ion{O}{iii}], and \ion{He}{ii} emission lines. In \S\ref{sec:models}, we explain our suite of ionization models computed using the {\tt MAPPINGS V} code \citep{Sutherland2017,mappingsV}, the fully empirical abundance patterns from \citet{Nicholls2017} and dust depletion scheme from \citet{Jenkins2009,Jenkins2014}, and the latest atomic data from {\tt CHIANTI} \citep{chianti10}. Our approach ensures that the only differences among our predictions are the ionization mechanisms.
We compile observations of nitrogen-excess galaxies from the literature, which we review in\S\ref{sec:data},  for comparison with these models. We use our new suite of models to assess new and existing emission line flux ratio diagrams to classify the nitrogen-excess galaxies at $z>5$ in \S\ref{sec:results} and find we can largely account for the apparent excess with shock excitation. Then, we discuss the implications for nitrogen abundances and high redshift ISM conditions in \S\ref{sec:discuss}, including corroborating evidence for shocks, reduction in measured nitrogen excess, and how Wolf Rayet stars may lead to the unique conditions we observe.

\section{A Library of Uniformly Treated Models}\label{sec:models}

Here, we present our new library of ionization models, which have uniformly implemented empirical abundance patterns and dust depletion treatment as well as updated atomic data. To date, no library of ionization models has been established on homogeneous foundations, which introduces possible systematic differences in both assumptions and approaches to numerical solutions and makes them unsuitable for this investigation. We take this homogeneous approach, drawing on empirical trends and physically-motivated inputs to build our library.
To ensure uniformity across our models, we make consistent use the \citet{Nicholls2017} empirical stellar abundance patterns scaled by the $\zeta_{\rm O}$ parameter (analogous to $Z/Z_\odot$) for $\zeta_{\rm O}=0.05$ to 2 (roughly 5 to 200\% $Z/Z_\odot$), the \citet{Jenkins2009,Jenkins2014} empirical dust depletion scheme with $F_\star=0.43$ (30\% of carbon, 80\% of silicon, and 97\% of iron are locked into dust grains) when dust is applicable, and the {\tt MAPPINGS V} code \citep[v5.2.1,][]{Sutherland2017,mappingsV} with atomic data from the {\tt CHIANTI} database \citep[v10,][]{chianti10} applied to all types models. For robustness in distinguishing between ionization sources, we consider photoionization by an AGN or by stellar populations and collisional ionization by shocks. We list the specific model parameters in Table \ref{tab:models}.

\begin{table*}
    \centering
    \caption{Model grids computed using {\tt MAPPINGS V} v5.2.1 \citep{Sutherland2017,mappingsV} with the {\tt CHIANTI} atomic data \citep{chianti10} using the \citet{Nicholls2017} empirical abundance patterns of $\zeta_{\rm O}=0.05$ to 2 (roughly 5 to 200\% $Z/Z_\odot$). We implement the \citet{Jenkins2009,Jenkins2014} empirical dust depletion scheme where dust is applicable, assuming $F_\star=0.43$ such that 30\% of carbon, 80\% of silicon, and 97\% of iron are locked into dust grains. For isochoric photoionization models (including the shock precursors) we cover a grid in density from 10 to $10^4$ \ccm\ (1 to $10^5$ \ccm\ in the case of shocks). For isobaric models, we cover a grid in $P/k$ from $10^5$ to $10^9\rm~K~cm^{-3}$.}
    \label{tab:models}
    \begin{tabular}{p{2in} l l l}
    Ionization Source & Ionization Strength & Parameters & Density Structure\\
    \hline
    {\tt BPASS} v2.2 \citep{bpass2.2}    &  $\rm\log U=-4$ to $-0.5$ & 3 to 7 Myr inst, 50 to 250 Myr cont &  isobaric or isochoric\\
    {\tt STARBURST99} \citep{starburst99v1,starburst99v2}    &  $\rm\log U=-4$ to $-0.5$ & 1 to 7 Myr inst, 50 to 300 Myr cont & isobaric or isochoric\\
    \citet{Jin2012} AGN SED     &  $\rm\log U=-4$ to $-1$ & $M_{\rm BH}=10^6$ to $10^7$ $\rm M_\odot$, $L/L_{\rm Edd}=0.01$, $0.3$ & isobaric\\
    \citet{optxagnf} {\tt OPTXAGNF} AGN SED    & $\rm\log U=-4$ to $-0.5$ & $M_{\rm BH}=10^4$ to $10^7$ $\rm M_\odot$, $L/L_{\rm Edd}=0.01$, $0.3$ & isobaric or isochoric \\
    radiative shock $+$& $v_{s}=50$ to $1000$ \kms & $B = 10^{-4}$, $0.1$, $1$, $10$ $\rm\mu G$ & isobaric\\
    \phantom{lipsum }bremmstrahlung precursor& & & isochoric \\
    dusty radiative shock $+$& $v_{s}=50$ to $200$ \kms & $B = 10^{-4}\rm~\mu G$ & isobaric\\
    \phantom{lipsum }bremmstrahlung precursor& & & isochoric \\    \end{tabular}
\end{table*}

\subsection{Photoionizaton}\label{sec:phionModels}

Perhaps the most fundamental assumed component of photoionization models is the ionizing photon spectral energy distribution (SED), which is the key distinction between stellar and non-stellar photoionization.

For SFGs, we consider two sets of model ionizing photon SEDs, assuming in both cases a \citet{Kroupa2001} IMF with a 120 $\rm M_\odot$ cutoff. For the first set, we use {\tt Starburst99} \citep{starburst99v1,starburst99v2,starburst99v2a}, selecting the Geneva isochrones accounting for stellar rotation \citep{Geneva1,Geneva2} and the WM-Basic \citep{WMBASIC} and CMFGEN \citep{CMFGEN1,CMFGEN2} atmospheres, interpolating over the predicted SEDs to approximately match the \cite{Nicholls2017} abundance patterns following \citet{Dopita2013}. To allow for evolution of extragalactic \ion{H}{ii} regions \citep[cf.][]{Dopita2005,Dopita2006a}, we choose single populations with ages of 1, 3, 5, 7, and 10 Myr as well as a continuous star formation scenario with 50, 100, 150, 200, and 300 Myr. For the second set, we use {\tt BPASS} \citep{bpass2.2}, which incorporates the effects of binary stars. Again accounting for possible evolution, we consider single populations with ages of 3, 5, and 7 Myr and continuous star formation with 50 to 250 Myr durations in 50 Myr intervals.

For AGN, we consider a disk-powerlaw SED as predicted by the {\tt OPTX AGNF} code \citep{optxagnf} and the \citet{Jin2012} multicolored blackbody with cool and hot inverse-Compton scattering components. These SEDs are parameterized by the black hole mass ($10^4$ to $10^7$ $M_\odot$) and accretion rate as a fraction of Eddington luminosity ($0.01$ and $0.3$) and include a non-disk (i.e., non-thermal) power law component ($L\nu\propto\nu^{-\alpha}$) with an index of $\alpha=2.2$ \citep{Dopita2014,optxagnf}.

Following previous investigations \citep[e.g.,][]{Dopita2002,Groves2004,Davies2016,optxagnf,Zhu2023}, we incorporate full treatment of dust and radiation pressure assuming an isobaric density profile for all photoionization models, which can well-approximate conditions in AGN and SFGs with complex gas geometries \citep[cf.][]{Dopita2002,Groves2004,Kewley2019}. For SFGs, we also consider an isochoric density profile, which may be justified in the cases of larger and/or more evolved \ion{H}{ii} regions ionized by stars \citep[e.g.,][]{Phillips2008}. For reference and illustrative purposes, we also include isochoric density profiles for a subset of our AGN models, although the effects on our proposed diagnostics are marginal \citep[cf. comparison of AGN models in][]{Zhu2023}. We note that some AGN narrow-line regions (NLRs) can be described by a powerlaw density profile \citep[e.g.,][ Flury et al. in prep]{Binette2024}; however, this scenario is beyond the current scope of our investigation and should not impact the proposed diagnostic regimes since, even at high initial densities, the power law profile line predictions are still consistent with the low-density limit for collisionally excited lines like \ion{O}{iii}] and \ion{N}{iii}] \citep[e.g.,][ Flury et al. in prep]{Binette2024}.

\begin{figure*}
    \centering
    \includegraphics[clip=true,trim={0 0.2in 0 0.15 in},width=\linewidth]{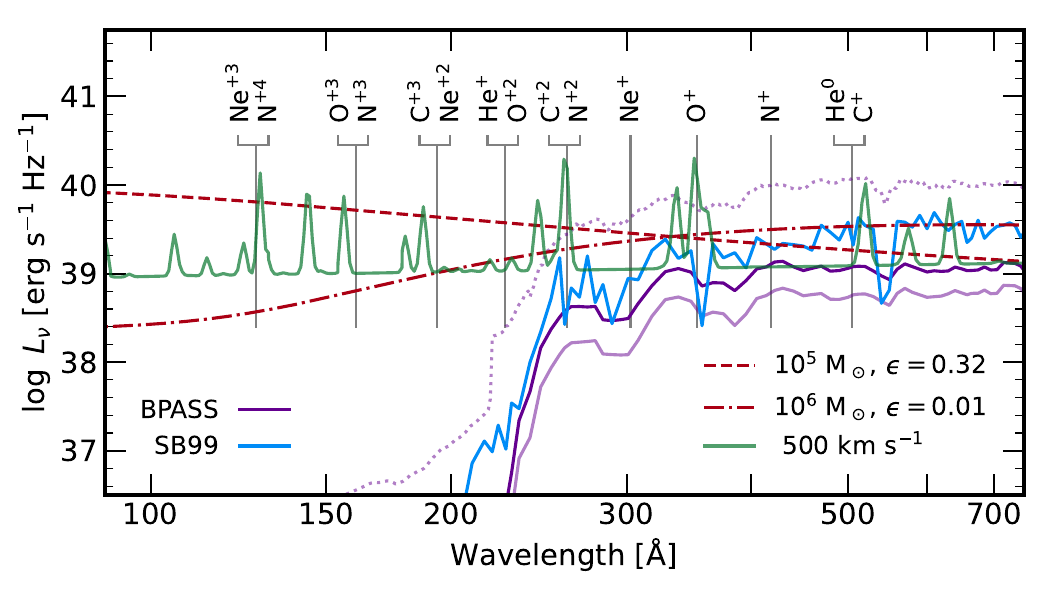}
    \caption{Ionizing photon spectral energy distribution models for {\tt oxaf} disk-powerlaw AGN (\citealt{optxagnf}, red), stellar populations (blue and purple), and a shock precursor (green). AGN models here indicate combinations of MBH mass and Eddington accretion rate. Stellar populations include instantaneous (solid blue and purple lines) and continuous (dotted purple line) for {\tt Starburst99} (\citealt{starburst99v1,starburst99v2,starburst99v2a}, blue) and {\tt BPASS} (\citealt{bpass2.2}, purple). {\tt MAPPINGS V} upstream ionizing spectrum consisting of bremsstrahlung continuum plus X-ray cooling lines shown in green for a 500 \kms\ shock, scaled for visibility.
    Ionization potential energies of various ions are labeled and indicated by grey lines (see Table \ref{tab:ionPot}).
    The difference in ionizing photon SED becomes most pronounced blueward of 228 \AA, corresponding to the ionization potentials for He$^{+}$ and O$^{+2}$. The nearby C$^{+2}$ and N$^{+2}$ ionization potentials at 263 \AA\ make these ions sensitive to the differences as well due to their photoionization cross-sections.}
    \label{fig:PhotIonSEDs}
\end{figure*}

To illustrate the power of emission lines of various ions in diagnosing photoionization excitation mechanisms, we compare ionizing photon SEDs from a subset of AGN and SF models in Figure \ref{fig:PhotIonSEDs} with notable ionization potential energies listed in Table \ref{tab:ionPot}. While similarities between the AGN and SFG ionizing photon SED shapes exist at wavelengths higher than $\sim228$ \AA, they diverge dramatically below this threshold. This difference arises from the \ion{He}{i} photoionization optical depth in stellar atmospheres and the Wein tail of the stellar blackbody at $T\sim35$ kK, effects which are not observed in characteristic AGN continua \citep[e.g.,][]{Jin2012,Stevans2014}. Thus, nebular emission from species with ionization potentials greater than the energy corresponding to the $\rm He^{+}$ edge can readily distinguish the type of photoionizing source. As a result, \ion{He}{ii} is an excellent discriminator between AGN and SFGs when paired with, e.g., \ion{O}{iii}]. Ions with photoionization cross sections which are still relatively large at this threshold will also differentiate between these excitation mechanisms. Under this criterion, \ion{C}{iv} or \ion{N}{iv}] can be sensitive diagnostics when paired with \ion{C}{iii}] or \ion{N}{iii}] since C$^{+2}$ and N$^{+2}$ share an ionization potential and have substantial photoionization cross sections below 228 \AA\ \citep[see, e.g.,][]{verner1996}.

\subsection{Shocks}\label{sec:shockModels}

Compact star-forming galaxies like those at high redshift are known to exhibit high-ionization shock signatures such as [\ion{O}{iv}] \W$25.89\rm~\mu m$ and [\ion{Ne}{v}] \W$3426$ \citep[e.g.,][]{Izotov2021}. Ambiguously classified galaxies in the BPT can also exhibit low-ionization shock-sensitive features like [\ion{S}{ii}] \WW$4068,74$ (e.g., Einsig et al. in prep, \citealt[e.g.,][]{Dopita2015}) and [\ion{O}{1}] \W6300 \citep[e.g.,][]{Allen2008,Dopita2015} which can distinguish shock ionization from other excitation mechanisms. Shocks may significantly affect the conditions in nebulae, including ionization structure and gas density, and have been reported in AGN \citep[e.g.,][ Flury et al. in prep]{Dopita1995a,Dopita2015,Dors2021,Riffel2021}, galactic winds \citep[e.g.,][]{Ho2014,Medling2015}, mergers \citep[e.g.,][]{Rich2011,Rich2015}, and supernovae \citep[e.g.,][]{Dopita2016}.

Within {\tt MAPPINGS V}, the shock model is treated as a time-dependent flow in the frame of the shock \citep[e.g.,][in particular \citealt{Sutherland1993} Fig 1-3 \& Table 4, \citealt{Allen2008} Fig 3 and \S4, \citealt{Sutherland2017} Fig 6]{Binette1985}, which we summarize here. The gas upstream of the shock has a given composition, ionization balance, density, and temperature. This gas  (called the ``precursor'') then flows downstream into a discontinuity described by the Rankine-Hugoniot jump conditions, the so-called shock front, and characterized by a shock velocity $v_s$ and magnetic field strength $B$. After entering the shock front, the gas rapidly reaches collisional ionization equilibrium at very high temperatures ($T_{s}\propto v_s^{2}$ from conservation of energy, with $T_s \ga 10^5$ K for $v_s\ga100$ \kms, e.g., \citealt{Binette1985,Draine1993,Sutherland2017}) with exact conditions dependent on the pre-shock gas properties. The hot gas cools via bremsstrahlung radiation and line emission in the soft X-rays \citep[e.g.,][]{Allen2008,Sutherland2017} in is approximately isobaric \citep[e.g.,][]{Draine1993,Sutherland1993}. After bremsstrahlung radiation sufficiently reduces the gas temperature below $\sim10^6\rm~K$, additional line emission mechanisms increase the cooling efficiency to rates much higher than the recombination rate, leading to a rapid (potentially catastrophic, e.g., \citealt{Falle1975,Falle1981,Smith1989}) decline in temperature. A consequence of this rapid cooling is that high ionization species like Ne$^{+4}$, produced by ionization energies $\geq$97.190 eV ($\sim T=10^6\rm~K$), persist in the radiative regime even as the gas approaches $10^4\rm~K$. Once the cooled ionized gas begins to recombine, the bremsstrahlung radiation from the hot gas within the shock (now downstream of the discontinuity) produces and sustains a photoionized region out to an ionization front which trails the shock front and which can produce secondary \citep[albeit radiatively negligible, e.g.,][]{Sutherland1993} shocks.

In addition to the flow solution, shock models must be treated self-consistently, i.e., the ionizing spectrum produced by bremsstrahlung radiation from the hot gas within the shock travels upstream into the pre-shock gas conditions since these conditions determine properties of the gas immediately after the shock front \citep[e.g.,][]{Shull1979,Sutherland1993,Draine1993,Dopita1996}. At $v_s\ga40$ \kms, heating of the pre-shock gas becomes important while ionization of the pre-shock gas becomes important at $v_s\ga80$ \kms\  \citep[][their Figures 4-5]{Sutherland2017}. In cases of high shock velocity ($v_s\ga200$ \kms), the pre-shock gas can be sufficiently ionized by the emergent bremsstrahlung radiation that it reaches photoionizaiton equilibrium, behaving like an \ion{H}{ii} region and even dominating the emergent nebular line emission \citep[e.g.,][]{Binette1985,Allen2008,Sutherland2017}. As such, incorporating these pre-shock conditions into the shock is essential.
To ensure self-consistency, we adopt the iterative shock-precursor scheme introduced for {\tt MAPPINGS} by \citet{Allen2008}, improved upon \citet{Sutherland2017}, and further discussed in \citet{Alarie2019} to account for effects of the heating and ionization of the precursor on the subsequent shock. The approach is implicit to {\tt MAPPINGS V} with dramatic improvements to, among other things, treatment of the shock ionizing spectrum, precursor structure, and radiative cooling \citep{Sutherland2017}. {\tt MAPPINGS V} computes an initial isobaric shock model under the Rankine-Hugoniot jump conditions, predicting an emergent ionizing spectrum, the ``upstream'' ionizing SED (see Figure \ref{fig:PhotIonSEDs}), which the code uses to compute a photoionization model for an isochoric one-sided slab \citep{Sutherland1993,Sutherland2017,Allen2008}, the ``precursor'', to obtain the pre-shock ionization balance and gas temperature. {\tt MAPPINGS V} recomputes the shock models assuming the precursor ionization balance and temperature for the initial conditions, generating a new upstream SED for the precursor photoionization model. We enforce a minimum of 5 iterations of this calculation as suggested in the literature \citep{Allen2008,Sutherland2017,Alarie2019} until the preshock temperature and ionization balance converge.

Due to the extreme temperature of the gas immediately after the shock front, the dust grains can be destroyed--via thermal (gas-grain collisions) and non-thermal (betatron acceleration) sputtering for smaller grains or shattering (grain-grain collisions) for larger grains--before the shocked gas enters the radiative phase \citep[e.g.,][]{Draine1993,Borkowski1995,Jones1996}. As such, we assume all of the dust grains are destroyed by the shock and thus do not consider the effects of dust in the majority of our shock models, a common convention in the literature \citep[e.g.,][]{Allen2008,Sutherland2017,Alarie2019}. To demonstrate possible effects of dust, we compute an additional set of shock models under conditions conducive to dust survival: (i) low gas temperatures after the shock front ($v_s\leq200$ \kms, recalling $T_s\propto v_s^2$) to mimic conditions limiting thermal sputtering and grain shattering plus (ii) no magnetic field ($B=0\mu$G), which eliminates non-thermal sputtering.

We compute shock models across a grid of velocities $v_s$ (50 to 1,000 \kms), densities (1 to $10^5$ \ccm), and magnetic field strengths ($10^{-4}$ to 10 $\mu$G). To obtain the predicted shock+precursor fluxes, we take the sum of the predicted shock and precursor line fluxes, weighted by the predicted H$\beta$ flux to account for the relative contributions of each component to the emergent spectrum.

\section{Ancillary Observations}\label{sec:data}

The aim of this study is to characterize the nitrogen excess recently reported in galaxies at $z\ga5$ using our library of ionization models based on homogeneous assumptions and implementations instead of drawing on novel physical explanations. We assemble a sample of nitrogen-excess galaxies from the literature. For context, we complement these high-redshift \emph{JWST} results with nearby ($z<0.1$) objects and a $z\sim2$ lensed galaxy, all with \ion{N}{iv}] and/or \ion{N}{iii}] emission reportedly due to nitrogen excess.


Several galaxies exhibiting nitrogen excess at high redshift are introduced in \S\ref{sec:intro}, including
GHZ2 \citep[$z=12.4$,][]{Castellano2024},
GN-z11 \citep[$z=10.6$, 3 pixel extraction, high resolution gratings][]{Bunker2023},
CEERS-1019 \citep[$z=8.678$,][]{MarquesChaves2024}, and
GS 3073 \citep[$z=5.55$, $0.3\times0.3^{\prime\prime}$ PRISM,][]{Ji2024}.
In addition to these sources, we consider the galaxies
RXCJ2248-ID \citep[$z=6.2$,][]{Topping2024a},
GS-z9-0 \citep[$z=9.432$, medium resolution grating,][]{Curti2024}, and
A173-zd6 \citep[$z=7.0435$,][]{Topping2024b}.
For these seven sources, we compile UV nebular line fluxes for \ion{N}{iv}] \WW1483,86, \ion{C}{iv} \WW1548,50, \ion{He}{ii} \W1640, \ion{O}{iii}] \WW1660,66, \ion{N}{iii}] \WW1747-54, and \ion{C}{iii}] \WW1907,09 from the literature when reported.

For a lower-redshift complement, we include three galaxies with properties similar to those exhibiting nitrogen excess at $z>5$. The VLT/MUSE observations of the $z=2.329$ the Lyman continuum (LyC) leaking knot of the Sunburst Arc indicate a nitrogen excess \citep{Vanzella2020,Pascale2023,Welch2024}. Green Pea analog Mrk 71 ($z=0.000805$) is quite similar to the Sunburst Arc in terms of its physical ISM properties and stellar populations \citep[e.g.,][]{Smith2023} and exhibits high temperatures and elevated \no\ relative to trends among \ion{H}{ii} regions \citep[e.g.,][]{Esteban2002,Esteban2009}. Nearby ($z=0.00547$) galaxy Mrk 996 has a reported nitrogen excess accompanied by strong Wolf-Rayet features indicative of enrichment in action \citep{Thuan1996,Thuan2008,James2009,Telles2014}. This blue dwarf compact star-forming galaxy has been considered as an analog to nitrogen-excess galaxies at the epoch of reionization \citep{Senchyna2023,MarquesChaves2024} and may help to provide local insight into the nature of the higher redshift populations. We use the recent, more sensitive \emph{HST}/COS observations of Mrk 996 from the COS Legacy Archive Spectroscopic SurveY \citep[CLASSY][]{Berg2022,Mingozzi2022}. Together, Mrk 71, Mrk 996, and the Sunburst Arc provide our lower redshift complement the nitrogen-loud galaxies at $z>5$. We also include 5 extreme ionization sources at $z\sim0$ analogous to high-redshift \ion{C}{iv} emitters from \citet{Jung2024} as these galaxies purportedly exhibit some contribution from shocks and are similar to the shock-excited [\ion{Ne}{v}] \W3426 emitters from \citet{Izotov2021}. To provide context for these extreme galaxies, we compare these extreme objects with a more complete sample of nearby ($z\sim0$) SFGs from CLASSY \citep[][]{Berg2022,Mingozzi2022} as a reference and foundation for our diagnostics.

\section{Results}\label{sec:results}

\begin{table}
    \caption{Lower and upper ionization potential energies for ions traced by the emission represented in our emission line flux ratio diagnostics. Columns indicate (i) the ion, (ii) the vacuum wavelength(s) of the associated emission line(s), (iii) the energy $E_{i-1\to i}$ necessary to produce the ion $N^{+1}$ by removing an electron from the lower ion species $N^{+i-1}$, and (iv) the energy $E_{i\to i+1}$ necessary to ionize $N^{+i}$ to the higher ion species $N^{+i+1}$. In the case of $\rm He^{+2}$, no electrons remain, and the species cannot be further ionized. {Here and throughout, the \ion{N}{iii}] \WW1747,49,50,52,54 quintuplet is expressed as \WW1747-54.} Values obtained from {\tt NIST} \citep{NIST_ASD}.}
    \label{tab:ionPot}
    \centering
    \begin{tabular}{rcrr}
    ion     & $\lambda$  & $E_{{i-1}\to {i}}$ & $E_{{i}\to {i+1}}$ \\
    & [\AA] & [eV] & [eV] \\
    \hline
$\rm He^{+2}$    & 1640    & 54.418 &  \\
$\rm C^{+2}$     & 1548,50 & 24.383 & 47.888 \\
$\rm C^{+3}$     & 1907,09 & 47.888 & 64.494 \\
$\rm N^{+2}$     & 1747-54   & 29.601 & 47.445 \\
$\rm N^{+3}$     & 1483,86 & 47.445 & 77.474 \\
$\rm O^{+\ph}$   & 3726,29 & 13.618 & 35.121 \\
$\rm O^{+2}$     & 1660,66 & 35.121 & 54.936 \\
$\rm Ne^{+2}$    & 3869    & 40.963 & 63.423 \\
$\rm Ne^{+3}$    & 2424    & 63.423 & 97.190 \\
$\rm Ne^{+4}$    & 3426    & 97.190 & 126.247 \\

    \end{tabular}

\end{table}

\begin{table}
    \caption{Sensitivity of various emission lines to electron density. Columns indicate (i) line species, (ii) the vacuum wavelength(s), (iii) the upper state of the electronic transition producing the line via spontaneous emission, and (iv) critical densities $n_{c}$ of the upper state. We calculate $n_{c}$ using {\tt PyNeb} \citep{pyneb} assuming $T_e\sim10^4$ K. Increasing $T_e$ results in a slight increase in $n_{c}$.}
    \label{tab:critDens}
    \centering
    \begin{tabular}{c c c c}
    species & $\lambda$    &  upper state & $n_{c}$\\
    & $[$\AA$]$ & & $\rm~[cm^{-3}]$ \\
    \hline
    {[}\ion{N}{iv}$]$       & 1483     & $\rm^3P_2$ & $10^5$ \\
    \phantom{[}\ion{N}{iv}$]$       & 1487     & $\rm^3P_1$ & $4\times10^9$ \\
    \ion{C}{iv}        & 1548,50  & $\rm^2P_{~}$ & $10^{15}$ \\
    \phantom{[}\ion{O}{iii}$]$     & 1660,66  & $\rm^5S_2$ & $4\times10^{10}$ \\
    \phantom{[}\ion{N}{iii}$]$        & 1747-54     & $\rm^4P_{~}$ & $10^{10}$ \\
    {[}\ion{C}{iii}$]$     & 1907     & $\rm^3P_2$ & $7\times10^4$ \\
    \phantom{[}\ion{C}{iii}$]$     & 1909     & $\rm^3P_1$ & $10^9$ \\
    $[$\ion{O}{ii}$]$  & 3726     & $\rm^2D_{\frac3 2}$ & $8\times10^3$ \\
    $[$\ion{O}{ii}$]$  & 3729     & $\rm^2D_{\frac5 2}$ & $3\times10^3$ \\
    $[$\ion{Ne}{iii}$]$ & 3869     & $\rm^1D_2$ & $10^7$
    \end{tabular}
\end{table}

We consider two classes of emission lines based on the critical densities of the upper states of their transitions, which we have determined using {\tt PyNeb} \citep{pyneb} and listed in Table \ref{tab:critDens}.
i) those which have high critical densities, including \ion{C}{iv} \WW1548,50, \ion{He}{ii}$\lambda1640$, \ion{O}{iii}] \WW$1660,6$, \ion{N}{iii}] \WW1747-54 quintuplet, and [\ion{Ne}{iii}] \W$3869$, making them excellent ionization, temperature, and abundance diagnostics,
and ii) those which have low critical densities, including the [\ion{N}{iv}] \W1483 and [\ion{C}{iii}] \W1907 lines and the [\ion{O}{ii}] \WW3726,9 doublet, making them excellent probes of gas density and shock excitation, particularly \ion{N}{iv}] and \ion{C}{iii}] in the cases where post-shock densities can be relatively high ($\ga10^4-10^5$ \ccm). We further consider in the construction of our diagnostics the ionizing energy of the ions associated with each of the emission lines, which we list in Table \ref{tab:ionPot}. The \ion{C}{iv} \WW1548,50 doublet, while promising due to its ionization potential and high critical density, can suffer from stellar contamination \citep[as in Mrk 71, e.g.,][]{Smith2023}, which may complicate its use as a nebular diagnostic \citep[e.g.,][]{Mingozzi2024}.

Since our aim is to understand the ionization mechanisms and related nitrogen abundance in galaxies at $z>5$, we limit our considered predictions to those for models of $\zeta_{\rm O}\leq0.5$ (\oh$\leq$8.35) to avoid accounting for observed flux ratios with unrealistically high metallicities. We also confirm that our models predict [\ion{O}{iii}] \W5007 / H$\beta$, [\ion{N}{ii}] \W6584 / H$\alpha$, [\ion{S}{ii}] \WW6716,31 / H$\alpha$, and [\ion{O}{i}] \W6300 / H$\alpha$ flux ratios within the typical range of \citet{Veillexu1987} BPT diagnostics.


\subsection{Previous Diagnostics for $z>5$}\label{sec:prevDiags}

\subsubsection{\ion{C}{iii}], \ion{C}{iv}, and \ion{He}{ii}}

\begin{figure}
    \centering
    \includegraphics[clip=True,trim={0 0.55in 0 0},width=\columnwidth]{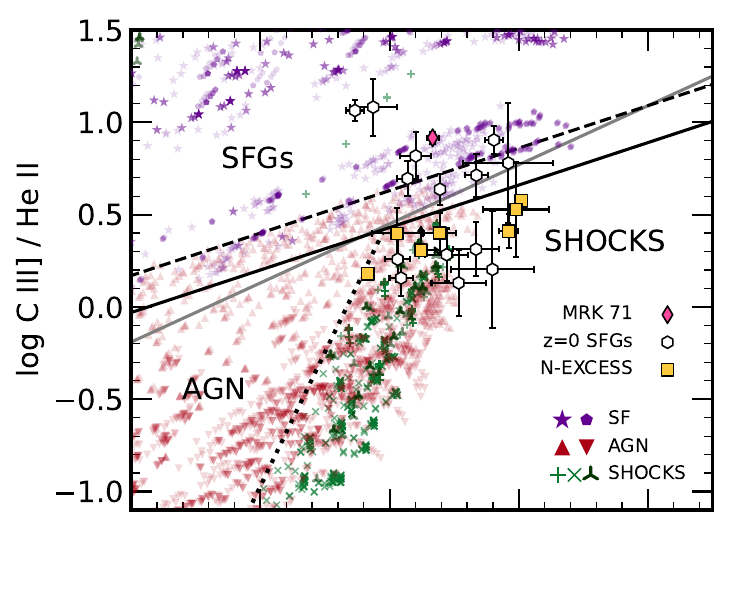}

    \includegraphics[clip=True,trim={0 0in 0 0.15in},width=\columnwidth]{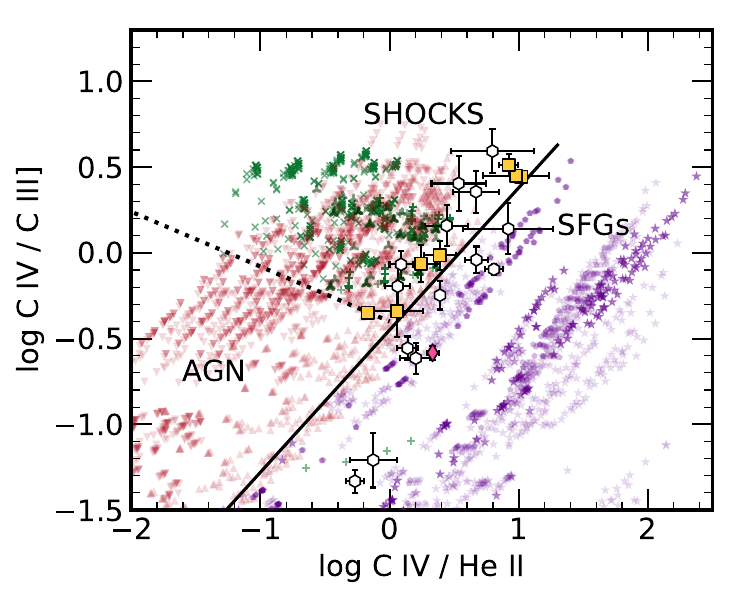}
    \caption{BPT-style diagnostics for \ion{C}{iii}]/\ion{He}{ii} (\textit{top}) and \ion{C}{iv}/\ion{C}{iii}] (\textit{bottom}) vs  \ion{C}{iv}/\ion{He}{ii}. Our model predictions for $\zeta_{\rm O}\leq0.5$ ($\la$50\% $Z_\odot$) are shown in purple for SFGs (stars and pentagons for instantaneous and continuous star formation, respectively), red for AGN (up and down triangles for the \citealt{Jin2012} and \citealt{optxagnf} SED models, respectively), and green for shocks (crosses and exes for $v_s\leq200$ \kms\ and $v_s>200$ \kms, the ``low'' and ``high'' shock velocity regimes, respectively; triradii for dusty shocks with $v_{s}=50$-$200$ \kms).
    Yellow squares correspond to nitrogen-excess galaxies at $z>5$ observed with \emph{JWST}. Pink diamond indicates the giant extragalactic \ion{H}{ii} region Mrk 71-A, a Green Pea analog exhibiting density enhancement and signatures of massive stars. White hexagons indicate the $z\sim0$ star-forming galaxies from CLASSY \citep[][]{Berg2022,Mingozzi2022} and from \citet{Jung2024}.  
    Grey line is the 10\% shock mixture boundary given by \citet{Jaskot2016}. Black lines indicate our limits for SFGs (solid), AGN (dashed), and shocks (dotted) as given by Equations \ref{eqn:c3he2c4he2} (\textit{top}) and \ref{eqn:c4c3c4he2} (\textit{bottom}).}
    \label{fig:c3he2c4he2}
\end{figure}

First proposed by \citet{VillarMartin1997}, diagnostics based on \ion{C}{iii}], \ion{C}{iv}, and \ion{He}{ii} rely predominantly on the differences in the $\rm C^{+}$, $\rm C^{+2}$, and $\rm He^{+}$ ionization potentials to distinguish between AGN and SFGs using \ion{C}{iii}]/\ion{He}{ii}, although \citet{VillarMartin1997} also demonstrate possible overlap between shocks and AGN. We show their primary diagnostic in the top panel of Figure \ref{fig:c3he2c4he2}. The much closer ionization potentials of $\rm C^{+2}$ and $\rm He^{+}$ cause \ion{C}{iv}/\ion{He}{ii} to provide some but less diagnostic power (see \citealt{Feltre2016}, their Figure 14). With our more uniform and empirically-motivated models, we find that \ion{C}{iii}] / \ion{He}{ii} is the primary discriminator between AGN and SFGs, with \ion{C}{iv} / \ion{He}{ii} being largely insensitive to differences in photoionizing SED. We also find that \ion{C}{iii}] / \ion{He}{ii} can distinguish between shocks and photoionization by stars, in keeping with the \citet{Jaskot2016} investigation of shock contributions to \ion{C}{iii}] (grey line in Figure \ref{fig:c3he2c4he2}). From our models, we confirm that \ion{C}{iv} / \ion{He}{ii} can isolate radiative shocks from a subset of the AGN models, overlapping only with model predictions of high ionization parameter. We provide definitions below for the flux ratio regimes for each ionization mechanism:
\begin{equation}\label{eqn:c3he2c4he2}
\begin{split}
    x=&\log_{10}\left(\frac{\textrm{\ion{C}{iv}}}{\text{\ion{He}{ii}}}\right),~y=\log_{10}\left(\frac{\textrm{\ion{C}{iii}]}}{\text{\ion{He}{ii}}}\right)\\
    y =&
    \begin{cases}
        0.23x+0.45, & \text{shock below, BSFG above} \\
        0.23x+0.63, & \text{AGN below, CSFG above} \\
        1.45x+0.48, & \text{shock below} \\
    \end{cases}
\end{split}
\end{equation}
where BSFG corresponds to the instantaneous burst star-forming galaxy and CSFG corresponds to a continuously star-forming galaxy.
We show these demarcations in Figure \ref{fig:c3he2c4he2} as solid, dashed, and dotted lines, respectively, with the \citet{Jaskot2016} 10\% shock-SFG mixing sequence for reference.

As evidenced by their location in Figure \ref{fig:c3he2c4he2}, all seven nitrogen-excess galaxies at $z>5$ are consistent with high ionization parameter AGN or shocks, as are several of the CLASSY SFGs and all of the extreme ionization SFGs. The location of two low-metallicity CLASSY galaxies, J0337-0502 (SBS0335-052 E) and J0934+5514 (I Zw 18 NW), and four of the five extreme ionization SFGs below the shock and SFG limits in Equation \ref{eqn:c3he2c4he2} may indicate that shocks are a dominant mechanism. \citet{Mingozzi2024} find in their own UV diagnostics that the two low metallicity CLASSY galaxies are star-formation dominated with evidence for additional shock ionization, as do \citet{Jung2024} for the five \ion{C}{iv} emitters. Mrk 71, known to exhibit high gas densities and stellar winds, resides squarely in the SFG locus; however, low velocity shock models with $v_s=100$ and 150 \kms\ straddle the observed flux ratios.

When paired with \ion{C}{iii}], \ion{C}{iv} becomes a far stronger ionization diagnostic. This flux ratio is relatively insensitive to metallicity as it depends on emission from ions of the same element \citep[see, e.g.,][Figure 14]{Mingozzi2022}. Thus this flux ratio less ambiguously assesses the hardness of the ionizing spectrum. Since the ionizing SEDs of AGN and stellar populations tend to diverge blueward of the $\rm C^{+}$ ionization potential (see Figure \ref{fig:PhotIonSEDs}), the addition of \ion{He}{ii} can provide the leverage necessary to distinguish between excitation regimes. Pairing \ion{C}{iv}/\ion{C}{iii}] with \ion{C}{iv}/\ion{He}{ii} thus provides a clear distinction between AGN and SFG photoionization as depicted in the bottom panel of Figure \ref{fig:c3he2c4he2} \citep[see also][their Figure 10]{Zhu2023}, a distinction more pronounced than with \ion{C}{iii}] / \ion{He}{ii} vs \ion{C}{iv} / \ion{He}{ii} but without any additional lines. Shocks continue to occupy the same space as high ionization parameter AGN. We define the ionization regimes as
\begin{equation}\label{eqn:c4c3c4he2}
\begin{split}
    x=&\log_{10}\left(\frac{\textrm{\ion{C}{iv}}}{\text{\ion{He}{ii}}}\right),~y=\log_{10}\left(\frac{\textrm{\ion{C}{iv}}}{\textrm{\ion{C}{iii}]}}\right)\\
    y =&
    \begin{cases}
        \phantom{-}0.83x-0.45, & \text{SFG below, shock \& AGN above} \\
        -0.32x-0.40, & \text{\phantom{SF below,} shock above}
    \end{cases}
\end{split}
\end{equation}
We show these demarcations in the bottom panel of Figure \ref{fig:c3he2c4he2}.

As with \ion{C}{iii}] / \ion{He}{ii} vs \ion{C}{iv} / \ion{He}{ii}, the seven nitrogen excess galaxies at $z>5$ reside just within the boundaries for the shock excitation regime but may also be consistent with high ionization AGN (see Figure \ref{fig:c3he2c4he2}). Interestingly, \citet{Topping2024b} classified A1703-zd6 and RXCJ2248-ID as consistent with SFG models when using \ion{C}{iii}]/\ion{C}{iv} and previous photoionization models based on scaled-solar abundance patterns. Accounting for a more robust and updated abundance pattern set combined with empirical dust depletion, we find that \ion{C}{iii}], \ion{C}{iv} consistently places both A1703-zd6 and RXCJ2248-ID just above the maximal SFG demarcation, as shown in Figure \ref{eqn:c4c3c4he2} as well as in other diagnostics containing \ion{C}{iii}]/\ion{C}{iv} (e.g., Figure \ref{fig:c3he2o3he2}). Not only does this difference speak to the significance of using appropriate abundance patterns to define flux ratio diagnostic spaces but also indicates that these two galaxies' nebular emission may be contaminated or even dominated by shocks.

Again, CLASSY galaxies J0337-0502 and J0934+5514 and the same four shock-contaminated \citet{Jung2024} SFGs fall within the shock excitation regime, albeit rather close to the demarcation for SFGs. Although residing in the SFG regime, Mrk 71 is also close to the demarcation. Together, the \ion{C}{iv}, \ion{C}{iii}], \ion{O}{iii}], and \ion{He}{ii} emission from these seven local SFGs suggests that shocks could describe the line ratios of the nitrogen-bright galaxies at high redshift. Similar \ion{C}{iv} emitters at the epoch of reionization may be associated with shocks, likely indicating feedback from massive stars. 
That being said, a diagnostic more sensitive to shock excitation is necessary to distinguish whether AGN or shocks are responsible for the observed emission in nitrogen-excess galaxies.

\subsubsection{\ion{C}{iii}], \ion{C}{iv}, and \ion{He}{ii} with \ion{O}{iii}] }

\begin{figure}
    \centering
    \includegraphics[clip=True,trim={0 0.55in 0 0},width=\columnwidth]{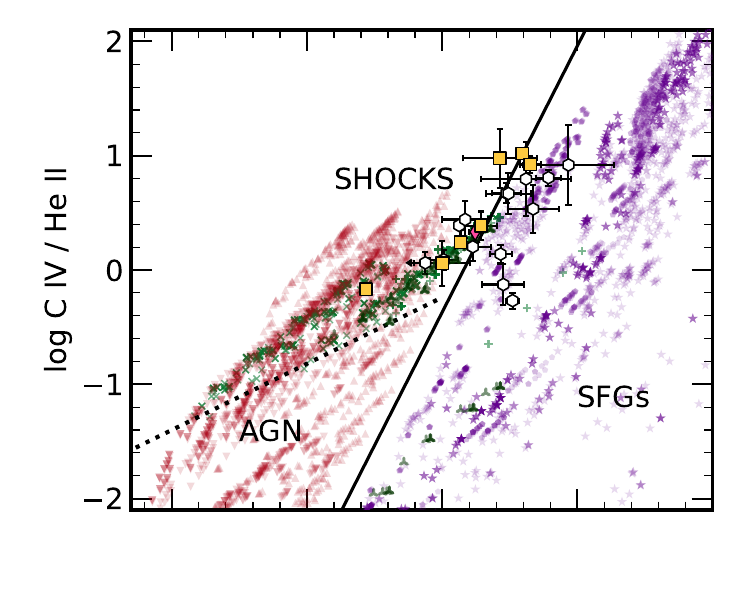}
    \includegraphics[clip=True,trim={0 0.55in 0 0.15in},width=\columnwidth]{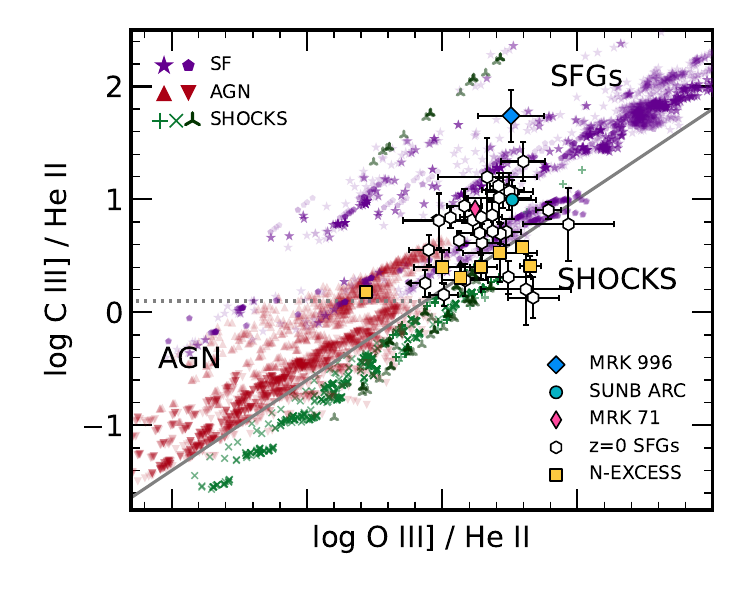}
    \includegraphics[clip=True,trim={0 0 0 0.1in},width=\columnwidth]{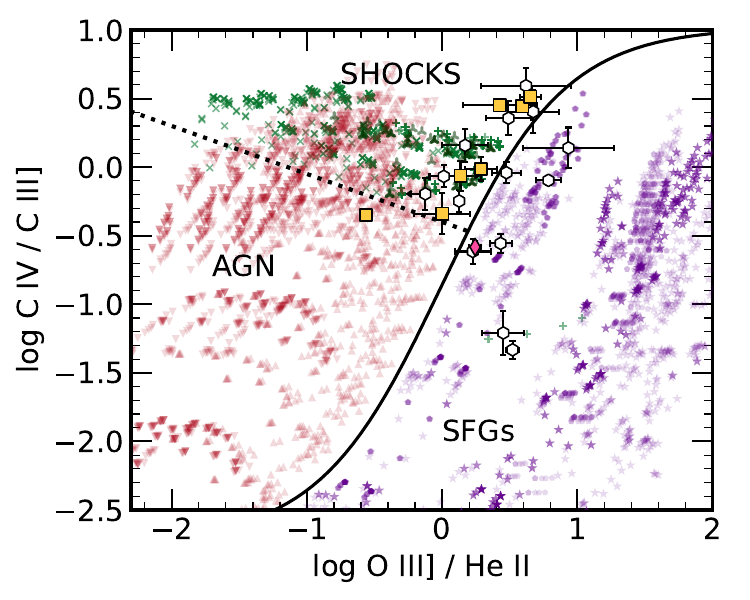}
    \caption{BPT-style diagnostic for \ion{C}{iv}/\ion{He}{ii} (\textit{top}), \ion{C}{iii}]/\ion{He}{ii} (\textit{center}), and \ion{C}{iv}/\ion{C}{iii}] (\textit{bottom}) vs  \ion{O}{iii}]/\ion{He}{ii}. Symbols as in Figure \ref{fig:c3he2c4he2}. Blue diamond indicates the low-redshift galaxy Mrk 996 exhibiting nitrogen excess and signatures of massive stars. Cyan circle indicates the nitrogen-excess knot of the Sunburst arc ($z=2.39$) exhibiting winds and Lyman continuum escape. Solid and dotted black lines in the top and bottom panels indicate demarcations given by Equations \ref{eqn:o3he2c4he2} and \ref{eqn:o3hec4c3}, respectively. Solid and dotten grey lines in the center panel indicate demarcations from \citet{Mingozzi2024}.}
    \label{fig:c3he2o3he2}
\end{figure}

The most fundamental shortcoming of the \ion{C}{iv}/\ion{He}{ii} and \ion{C}{iii}]/\ion{He}{ii} vs \ion{O}{iii}]/\ion{He}{ii} diagnostics is that, given both axes are normalized to \ion{He}{ii}, the flux ratios are extremely sensitive to metallicity \citep[the C/O ratio, e.g.,][]{Jaskot2016,Gutkin2016} and the preferential depletion of carbon into dust grains relative to oxygen \citep[e.g.,][]{Gutkin2016} in addition to ionization parameter. We find this sensitivity to be the case in our model predictions as well, limiting the extent to which these paired flux ratios can function as an ionization diagnostic.

For \ion{C}{iv}, the exceptionally high critical densities of the $\rm^2P$ states mitigates any effects of shocks on the line diagnostics. As a result, \ion{C}{iv}/\ion{He}{ii} is more sensitive to the precursor photoionized by the shock emission than \ion{C}{iii}]/\ion{He}{ii}. Given the similarities between the AGN and shock bremmstrahlung SEDs in Figure \ref{fig:PhotIonSEDs}, \ion{C}{iv} should not distinguish well between shocks and high ionization AGN, which is consistent with the results in Figure \ref{fig:c3he2o3he2}. As shock velocity decreases, so does the peak of the emergent ionizing SED, making the precursors of lower velocity shocks appear similar to high ionization SFGs, further diminishing the ability of \ion{C}{iv}/\ion{He}{ii} to distinguish shocks from other ionizing mechanisms. However, \ion{C}{iv} and \ion{He}{ii} together trace the stark differences in AGN and stellar SEDs between 200 and 300 \AA. Since \ion{O}{iii}] and \ion{He}{ii} are similarly sensitive to the drop ion stellar ionizing SEDs, it is perhaps no surprise that \ion{C}{iv}/\ion{He}{ii} vs \ion{O}{iii}]/\ion{He}{ii} provides one of the best demarcations between AGN and SFG photoionization among the diagnostics considered in the literature. We define demarcations between AGN and SFGs such that
\begin{equation}\label{eqn:o3he2c4he2}
\begin{split}
    x=&\log_{10}\left(\frac{\textrm{\ion{O}{iiii}]}}{\text{\ion{He}{ii}}}\right),~y=\log_{10}\left(\frac{\textrm{\ion{C}{iv}}}{\text{\ion{He}{ii}}}\right)\\
    y =&
    \begin{cases}
        2.33x-0.37, & \text{SFG below, AGN above} \\
        0.58x-0.24, & \text{\phantom{SFG below, }shock above} \\
    \end{cases}
\end{split}
\end{equation}
and
\begin{equation}\label{eqn:o3hec4c3}
\begin{split}
    x=&\log_{10}\left(\frac{\textrm{\ion{O}{iiii}]}}{\text{\ion{He}{ii}}}\right),~y=\log_{10}\left(\frac{\textrm{\ion{C}{iv}}}{\text{\ion{C}{iii}]}}\right)\\
    y =&
    \begin{cases}
        3.75\left(1+e^{-2.20x}\right)^{-1}-2.73, & \text{SFG below, shock \& AGN above} \\
        -0.35x-0.38, & \text{\phantom{SFG below, }shock above} \\
    \end{cases}
\end{split}
\end{equation}
which we show in the top and bottom panels of Figure \ref{fig:c3he2o3he2} with solid and dotted lines, respectively.

We find the effects of shock compression on the [\ion{C}{iii}] line can serve to further separate the shock models from those of AGN with high ionization parameter, a result of the four orders of magnitude difference in critical density between the \ion{O}{iii}] transitions and [\ion{C}{iii}] \W1907 (see Table \ref{tab:critDens}). 
Perhaps most compelling about this result is that the seven $z>5$ nitrogen-excess galaxies either reside in the shock locus of this diagnostic or within uncertainties could readily occupy it. The CLASSY SFGs and our nitrogen-excess analogs occupy a sequence extending from the shock regime into the heart of the SF locus, with uncertainties allowing for the majority of SFGs to occupy either ionization regime. While Mrk~996 does not appear to be shock excited according to the \citet{Mingozzi2024} demarcation, this galaxy, Mrk 71, and the Sunburst arc could be explained by lower density ($n_{\rm H}=1$ \ccm) shock models with $v_{s}=50$ to $100$ \kms. Alternatively, mixing of the SFG and shocks within the COS aperture could be pushing these galaxies toward the demarcation \citep[see shock-SF mixing explored for other diagnostics in][]{Jaskot2016}.

We note that the distinction between AGN, SFGs, and shocks only holds at low metallicity. Expanding to high metallicities causes the shock models to occupy a portion of the SFG locus where our local nitrogen-excess reference objects reside, as well as a significant portion of the CLASSY galaxies. Further complications include the substantial overlap between AGN and SFG model predictions. The putative AGN GS 3073 resides in an ambiguous part of the diagnostics, offset from other nitrogen-excess galaxies due to a relatively weak \ion{O}{iii}] emission relative to \ion{He}{ii}. While enhanced \ion{He}{ii} is a possible signature of AGN activity, we cannot rule out a scenario such as SFG-shock mixing, leaving confirmation of the presence of an AGN to other diagnostics.

In general, \ion{C}{iii}] / \ion{He}{ii} vs \ion{O}{iii}] / \ion{He}{ii} can differentiate between pure shocks and other mechanisms at low metallicity. We attribute this ability to collisional suppression of [\ion{C}{iii}] \W1907 and augmentation of \ion{O}{iii}] \WW1660,66. While \ion{He}{ii} is a strong discriminator of SFGs due to the He$^{+}$ ionization potential, sufficient similarities in AGN and SFG ionizing SEDs at the ionization potentials of O$^{+}$ and C$^{+}$ limit the diagnostic power of \ion{C}{iii}] \WW1907,09 and \ion{O}{iii}] \WW1660,66 for photoionization sources.

\subsubsection{[\ion{Ne}{iii}] and [\ion{O}{ii}]}

The [\ion{Ne}{iii}] \W3869 / [\ion{O}{ii}] \WW3726,9 ratio is purported to distinguish between various excitation mechanisms \citep[the so-called ``OHNO'' diagram,][]{Backhaus2022}. We have explored [\ion{Ne}{iii}]/[\ion{O}{ii}] and find that this flux ratio, while a good probe of ionization parameter \citep[see, e.g.,][]{Levesque2014}, does not discriminate between excitation mechanisms \citep[see similar results in][]{Zhu2023}. We attribute this lack of diagnostic power to the similar shapes of ionizing SEDs at these ionization potential energies (Figure \ref{fig:PhotIonSEDs}).

\subsection{A Low-Resolution Rest-UV Diagnostic for $z>5$}

Oft-employed emission line flux ratio diagnostics \citep[notably the BPT][]{BPT1981,Veillexu1987} are designed with two observational limitations in mind: (i) detectability of the line, including resolution from adjacent features, and (ii) sensitivity to extinction. As such, lines considered must not only meet the scientific objectives outlined above but also satisfy the observational restrictions. Namely, lines must be bright and close in wavelength to limit the \emph{relative} effects of extinction while being sufficiently separated to be resolved by conventional spectrographs. In the lattermost case, the low resolutions of \emph{JWST}/NIRSpec PRISM/CLEAR at $z>3$ and \emph{HST}/STIS G230L at $z<0.5$ can complicate the use of various diagnostics. Features like \ion{He}{ii} and \ion{O}{iii}] will often be blended and difficult to separate accurately, potentially introducing bias into how diagnostic diagrams are interpreted. Previously proposed diagnostics \citep[e.g.,][]{VillarMartin1997,Feltre2016,Jaskot2016,Mingozzi2024} assume resolution of the \ion{He}{ii} and \ion{O}{iii}] blend, that the 400 \AA\ separation between \ion{C}{iv} and \ion{C}{iii}] is insensitive to dust, or that the UV continuum is only described by the ionizing source -- as is the case for all equivalent width diagnostics -- despite the fact that stellar populations older than 10 Myr (i.e., non-ionizing) can still contribute appreciably to the FUV continuum without affecting the ionizing continuum \citep[e.g.,][]{Chisholm2019}.   

\begin{figure}
    \centering
    \includegraphics[width=\columnwidth]{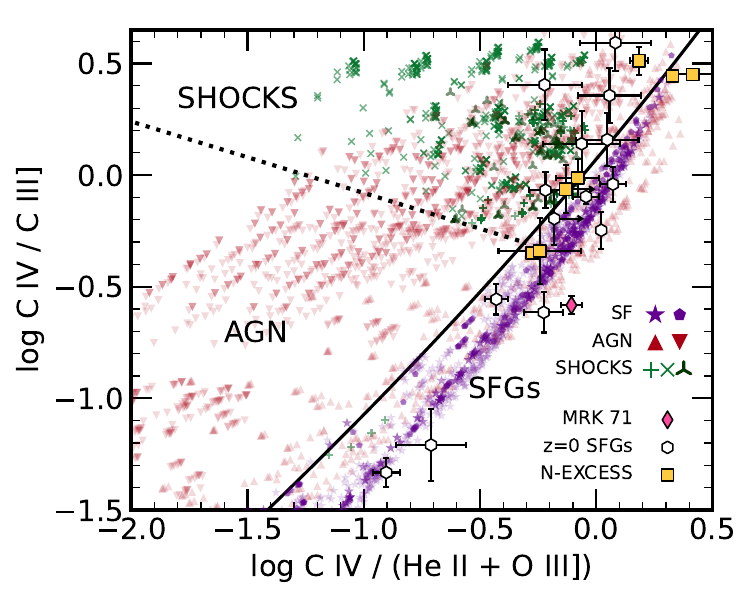}
    \caption{BPT-style diagnostics for \ion{C}{iv}/\ion{C}{iii}] vs  \ion{C}{iv}/(\ion{He}{ii}+\ion{O}{iii}]). Yellow squares correspond to nitrogen-excess galaxies at $z>5$ observed with \emph{JWST}. Symbols as in Figure \ref{fig:c3he2c4he2}. Solid, and dotted lines correspond to Equation \ref{eqn:c4c3c4o3he2} to distinguish excitation regimes, which we have labeled.}
    \label{fig:c4c3he2o3}
\end{figure}

With these considerations in mind, we propose a BPT-style flux ratio diagnostic that not only is motivated by previous diagnostics but also accounts for the low resolution of \emph{JWST}/NIRSpec PRISM/CLEAR. This diagnostic compares the bright \ion{C}{iii}] and \ion{C}{iv} lines with the \ion{He}{ii} + \ion{O}{iii}] blend
to distinguish different flux ratio regimes explicable by SFGs, AGN, and radiative shocks.
We define these regimes using maximum/minimum demarcations following previous sections such that
\begin{equation}\label{eqn:c4c3c4o3he2}
\begin{split}
    x=&\log_{10}\left(\frac{\textrm{\ion{C}{iv}]}}{\text{\ion{He}{ii}}+\text{\ion{O}{iii}]}}\right),~y=\log_{10}\left(\frac{\textrm{\ion{C}{iv}}}{\text{\ion{C}{iii}]}}\right)\\
    y =&
    \begin{cases}
        -\left(43.41-16.39x\right)^{1/2}+6.66, & \text{SF below, shock above} \\
        -0.32x+0.4, & \text{\phantom{SF below, }shock above} \\
    \end{cases}
\end{split}
\end{equation}
We show these demarcations in Figure \ref{fig:c4c3he2o3} as solid and dotted lines, respectively.

As evident in Figure \ref{fig:c4c3he2o3}, the \ion{C}{iii}]/(\ion{He}{ii}+\ion{O}{iii}]) ratio serves primarily to distinguish between SF and AGN photoionization, with a sharp upper limit in \ion{C}{iii}]/(\ion{He}{ii}+\ion{O}{iii}]) for the AGN model predictions. A similar lower limit in \ion{C}{iii}]/(\ion{He}{ii}+\ion{O}{iii}]) exists for the star forming galaxies. While these limits overlap by $\sim0.1$ to 0.3 dex, the AGN and SFG regions are otherwise distinctly separate. As with previous diagnostics considered, the nitrogen-excess galaxies reside within this overlap, making direct classification ambiguous at best. Additional diagnostics assuming higher resolution and higher sensitivity will therefore be necessary to ascertain the excitation mechanism(s) at play in the object of interest.

Shocks, while overlapping with the AGN models, occupy the upper left of Fig \ref{fig:c4c3he2o3}, owing primarily to collisional suppression of \ion{C}{iii}]. For a fixed set of parameters, varying the shock velocity from 100 to 1000 \kms\ leads to a decrease in \ion{C}{iii}]/(\ion{He}{ii}+\ion{O}{iii}]) but an increase in \ion{C}{iv}/\ion{C}{iii}].
This trend arises predominantly from the ionization structure of the shock+precursor: the number of $\rm He^{+1}$ ionizing photons increases by nearly three orders of magnitude from $v_{s}=100$ to 1000 \kms\ while those of $\rm O^{+}$ and $\rm C^{+}$ increase far less dramatically  \citep[cf.][]{Allen2008}, which augment the \ion{He}{ii} line substantially relative to the \ion{O}{iii}] and \ion{C}{iii}] doublets.
As result, the $z>5$ galaxies exhibiting nitrogen excess are most consistent with low-velocity shocks among our shock models. If shocks are indeed responsible (at least in part) for the ambiguous classification of these galaxies, then the shock velocities must be low. However, to distinguish readily between shocks, AGN, and SFGs, a more robust diagnostic is needed, one which may require higher resolution and sensitivity than is available with \emph{JWST}/NIRSpec PRISM/CLEAR or the \emph{HST}/COS or /STIS -L gratings.

\subsection{Insights from \ion{O}{iii}]/\ion{He}{ii} with \ion{N}{iii}] and \ion{N}{iv}]}

\begin{figure}
    \centering
    \includegraphics[clip=True,trim={0 0.55in 0 0},width=\columnwidth]{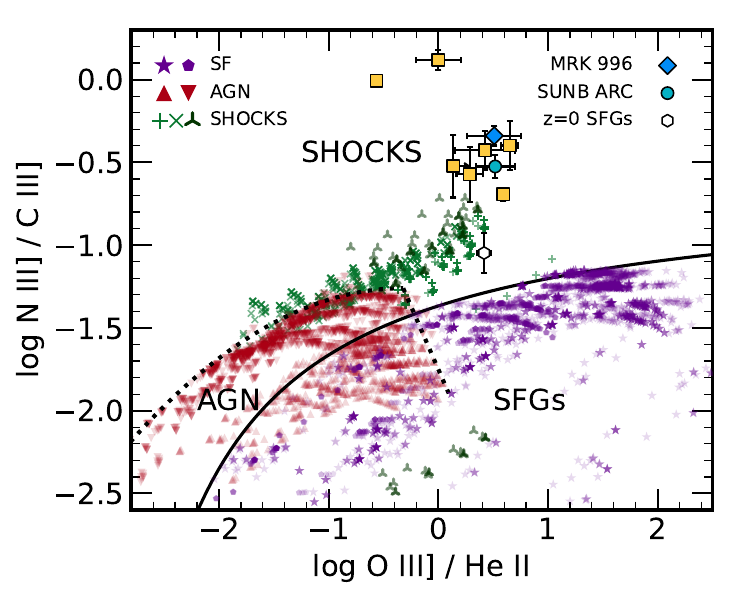}
    \includegraphics[clip=True,trim={0 0 0 0.15in},width=\columnwidth]{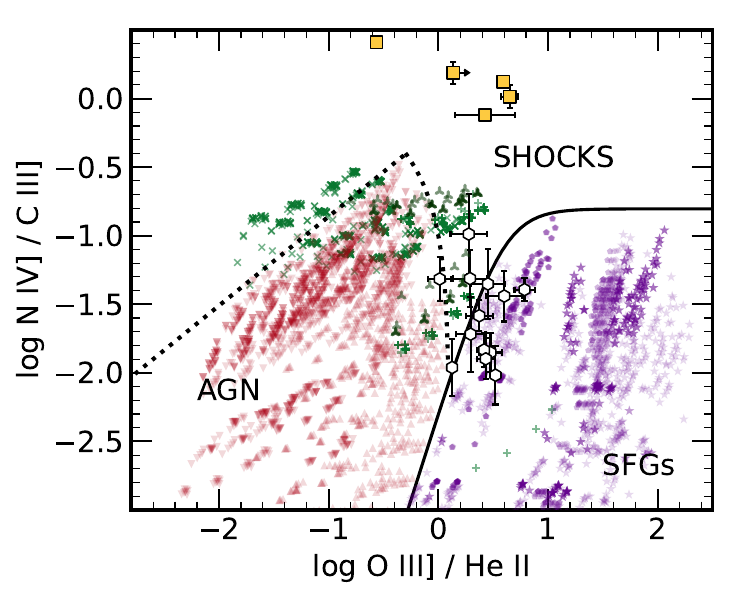}
    \caption{BPT-style diagnostics for \ion{N}{iii}]/\ion{C}{iii}] vs \ion{O}{iii}]/\ion{He}{ii} (\textit{top}) and \ion{N}{iv}]/\ion{C}{iii}] vs \ion{O}{iii}]/\ion{He}{ii} (\textit{bottom}). Symbols as in Figures \ref{fig:c3he2c4he2}-\ref{fig:c4c3he2o3}. Solid, dotted, and long-dashed lines correspond to Equations \ref{eqn:n3c3o3he2} and \ref{eqn:n4c3o3he2} in the top and bottom panels, respectively, to distinguish excitation regimes. 
    The CLASSY galaxy appearing as a filled white hexagon in both panels is J1253-3012 (SHOC391).
    Radiative shocks occupy a unique space where extreme densities and sufficiently extreme radiation can populate the diagram above the solid, dashed, and dotted lines. The location of the high redshift galaxies in these diagnostic diagrams suggests that radiative shocks are the dominant excitation mechanism for these objects and may account for some, if not all, of the apparent nitrogen excess.}
    \label{fig:n3c3o3he2}
\end{figure}

If higher resolution and sensitivity are attainable, more informative diagnostic lines and flux ratios can be considered. Certainly, resolving \ion{O}{iii}] from \ion{He}{ii} will provide considerable diagnostic leverage, as seen above, as will detecting the nitrogen lines linked both to various density regimes and to the apparent nitrogen excess. In Figure \ref{fig:n3c3o3he2}, we compare both \ion{N}{iii}]/\ion{C}{iii}] and \ion{N}{iv}] /\ion{C}{iii}] to \ion{O}{iii}] / \ion{He}{ii}.
The \ion{N}{iii}]/\ion{C}{iii}] and \ion{N}{iv}]/\ion{C}{iii}] ratios should be relatively insensitive to metallicity at \oh$\sim$7-8\ due to the enrichment pathways of nitrogen and carbon \citep[e.g.][ Arellano-Cordova et al. in prep]{Henry2000,Kobayashi2020,Curti2024}, making these flux ratios excellent tracers of various gas conditions. Given the critical densities of their respective transitions in Table \ref{tab:critDens}, \ion{N}{iii}]/\ion{C}{iii}] serves primarily as a diagnostic of dense gas. On the other hand, \ion{N}{iv}]/\ion{C}{iii}] serves primarily as a diagnostic of ionization because, from Table \ref{tab:ionPot}, $\rm N^{+3}$ has a similar ionization potential to $\rm C^{+3}$.
While \ion{O}{iii}]/\ion{He}{ii} is sensitive to metallicity, this ratio is also strongly dependent on ionization, tracing the extreme UV at 353 \AA\ and 228 \AA, respectively, where the ionizing capabilities of stellar populations are expected to decline precipitously while AGN and fast radiative shocks are not. Thus, \ion{O}{iii}]/\ion{He}{ii} is an excellent diagnostic for distinguishing between stellar and non-stellar ionizing sources.

In the top panel of Figure \ref{fig:n3c3o3he2}, we see a reduction in \ion{C}{iii}] flux relative to \ion{N}{iii}] due to shock enhancement of the gas density, causing shock models to reside in a unique flux ratio space. A similar effect is seen in the lower panel of Figure \ref{fig:n3c3o3he2}, albeit less pronounced given the combined effects of ionization and lower critical densities of the \ion{N}{iv}] transitions. The AGN and star-forming galaxy models additionally reside in largely unique flux ratio spaces, making these diagrams useful diagnostics for a variety of excitation sources. The velocity gradient noted earlier persists in these diagnostics, with increasing $v_S$ corresponding to decreasing \ion{O}{iii}]/\ion{He}{ii}, decreasing \ion{N}{iii}]/\ion{C}{iii}], and increasing \ion{N}{iv}]/\ion{C}{iii}].

To distinguish among the various excitation regimes in Figure \ref{fig:n3c3o3he2}, we provide the following quantitative diagnostics for flux ratios:

\begin{equation}\label{eqn:n3c3o3he2}
\begin{split}
    x=&\log_{10}\left(\frac{\textrm{\ion{O}{iii}]}}{\text{\ion{He}{ii}}}\right),~y=\log_{10}\left(\frac{\textrm{\ion{N}{iii}]}}{\textrm{\ion{C}{iii}]}}\right)\\
    y =&
    \begin{cases}
        \frac{-2.6}{x+3.5}-0.6, & \text{SFG below, shock above} \\
        -0.15(x-0.32)^2-1.26, & \text{AGN below, shock above} \\
        -1.53x-1.5, & \text{AGN below} \\
    \end{cases}
\end{split}
\end{equation}
and
\begin{equation}\label{eqn:n4c3o3he2}
\begin{split}
    x=&\log_{10}\left(\frac{\textrm{\ion{O}{iii}]}}{\text{\ion{He}{ii}}}\right),~y=\log_{10}\left(\frac{\textrm{\ion{N}{iv}]}}{\textrm{\ion{C}{iii}]}}\right) \\
    y =&
    \begin{cases}
        \log_{10}\left(12.6+10^{-2x+2.1}\right) & \text{SFG below, shock \& AGN above} \\
        -\log_{10}(x+0.10), & \text{AGN below, shock above} \\
        0.65x-0.21, & \text{AGN below, shock above} \\
    \end{cases}
\end{split}
\end{equation}
We show these demarcations in Figure \ref{fig:n3c3o3he2} as dotted, solid, and dashed lines, respectively. \ion{N}{iii}] is noticeably better at distinguishing shocks from AGN than \ion{N}{iv}], which we attribute to collisional suppression of the [\ion{N}{iv}] \W1483 line due to its comparatively low critical density.

In Figure \ref{fig:n3c3o3he2}, we also show observations of nitrogen-excess galaxies at $z>5$ observed with \emph{JWST} and compiled from the literature. Five galaxies, CEERS-1019, RXCJ2248-ID, GHZ2, GS-z9-0, and A1703-zd6, are offset from the low-velocity shock models by $\leq0.5$ dex, as are the analog galaxies Mrk 996 and the Sunburst arc. Rerunning selective cases of our shock models with a 0.5 dex increase in \no\ brings the dust-free models into agreement with the observations. Dusty shock models require only a 0.3~dex increase in \no\ to match these five galaxies due to effects such as radiative efficiency and depletion of C into graphitic grains. The \citet{Nicholls2017} abundance patterns for \no\ contain a scatter of $\sim0.3$ dex in the range of \oh\ comprising these galaxies and our models. Therefore, the nitrogen abundance ``excess" observed could instead be an increase in flux attributable to shocks combined with a bias toward detecting brighter nitrogen lines from slightly enhanced \no. The \no\ abundance in these galaxies would agree with \no\ vs \oh\ determined from Galactic stars and blue compact dwarf galaxies. Such excess could readily arise from enrichment by Wolf Rayet stars, as seen locally \citep[e.g.,][]{Bik2018,AbrilMelgarejo2024}. We discuss this possibility further in \S\ref{sec:discuss}.

One SFG from CLASSY, J1253-3012, appears in both diagrams in Figure \ref{fig:n3c3o3he2} and is known to exhibit both enhanced nitrogen (\oh$=$7.98, \no$=-$0.92, Arellano-Cordova in prep., cf. \citealt{Guseva2011}) and Wolf Rayet stars \citep[e.g.,][]{Brinchmann2008,Shirazi2012}. While this target was not initially considered as a possible nitrogen-excess analog, the properties of this galaxy indicate both shock excitation and high \no\ largely consistent with the scatter in local trends from \citet{Nicholls2017}. As with the five other shock-excited galaxies with nitrogen excess, the \no\ excess could arise from enrichment ``in action'' via Wolf Rayet stars.

The two outliers, GN-z11 and GS 3073, would still require a $\ga0.8$ dex excess in \no\ to reproduce the flux ratios using our shock models. Given possible AGN signatures in these objects \citep[][respectively]{Maiolino2024,Schaerer2024,Ji2024}, particulalry with the detected coronal lines in GS 3073, we propose a shock-AGN mixture where the AGN illuminates the precursor, drives a supersonic wind which shocks and compresses the gas, and prevents recombination by sustaining the post-shock radiative regime. Modeling such a scenario is beyond the scope of this paper; however, we hypothesize that the observed flux ratios arise from an AGN-shock mixture with \no\ values similar to those of the other nitrogen-excess galaxies considered here.



\section{Implications for Nitrogen}\label{sec:discuss}

\subsection{Substantiating Evidence For Shocks}

We have demonstrated above that, among the models considered in this work, shocks are the only excitation mechanisms which can account for the observed \ion{N}{iii}]/\ion{C}{iii}] and \ion{N}{iv}]/\ion{C}{iii}] flux ratios and are consistent with various \ion{C}{iii}], \ion{C}{iv}, and \ion{He}{ii} flux ratios. Here, we examine evidence to support this agreement. The shock models which best describe the observations of $z>5$ nitrogen-excess galaxies and their nearby analogs (Mrk 71, Mrk 996, and the Sunburst arc) in Figures \ref{fig:c4c3he2o3} and \ref{fig:n3c3o3he2} are those in the ``warm partially ionized'' regime described by \citet{Sutherland2017} with shock velocities $v_s\leq200$ \kms\ \citep[][]{Allen2008,Sutherland2017}. Such wind velocities are observed and readily achievable in star-forming galaxies across a broad swath of cosmic time \citep[e.g.,][]{Steidel2010,Heckman2011,Heckman2015} and can be attributed to outflows, stellar winds, and/or supernovae. Furthermore, winds in this velocity range are observed in cases of nitrogen excess at $z>10$ \citep[e.g., GHZ2,][]{Castellano2024} and at lower redshift \citep[e.g., in Mrk 996 and the Sunburst arc][]{James2009,RiveraThorsen2017,Vanzella2020,Mainali2022} and can occur even when feedback is suppressed \citep[e.g., in Mrk 71,][]{Oey2017}. Interestingly, CLASSY galaxy J1253-3012 (SHOC391) exhibits \ion{N}{iv}] and \ion{N}{iii}] in the shock regime in Figure \ref{fig:n3c3o3he2} as well as wind velocities between 90 and 140 \kms\ \citep[e.g.,][]{Xu2022}, making it an ideal candidate for contributions from low-velocity shocks. Both the emission line diagnostics presented in Figures \ref{fig:c4c3he2o3} and \ref{fig:n3c3o3he2} and the kinematic evidence point to shocks as possible drivers of the line excitation.

Two CLASSY galaxies, J0337-0502 (SBS0335-052 E) and J0934+5514 (I Zw 18 NW), appear in the shock region of the \ion{C}{iii}], \ion{C}{iv}, and \ion{He}{ii} diagrams discussed in \S\ref{sec:prevDiags}. Both SBS0335-052 E and I Zw 18 NW exhibit strong evidence for Wolf Rayet stars \citep[e.g.,][]{Thuan1997,Izotov1997,Izotov2006} and gas temperatures in excess of 20 kK \citep[e.g.,][]{Izotov2006}.
The northern knot of I Zw 18 contains a substantial wind of ionized gas with velocities of 60-80 \kms\ \citep[e.g.,][]{Petrosian1997,ArroyoPolonio2024}, capable of driving low-velocity shocks consistent with its \ion{C}{iii}], \ion{C}{iv}, \ion{He}{ii} flux ratios. I Zw 18 also exhibits a small amount of ionized gas in a superwind reaching as high as $3\times10^3$ \kms\ \citep{ArroyoPolonio2024}, which may indicate suppressed stellar feedback such as in Mrk 71 \citep[e.g.,][]{Oey2017}. SBS0335-052 E exhibits similar low-velocity shocks of $\sim100$ \kms\ in its older star-forming knots and much slower winds of 30-40 \kms\ in its youngest star-forming knots \citep[e.g.,][]{Izotov2006}. The slightly older star-forming regions have probably undergone supernova \citep{Izotov2006}, which could account for any low-velocity shock emission. The younger star-forming regions are likely associated with stellar winds, which would impart significantly less mechanical energy and thus drive shocks in the ``warm neutral'' regime of \citet{Sutherland2017} with photoionization by the earliest stars dominating the nebular emission.

\subsection{Shock Heating}

As discussed in \S\ref{sec:shockModels}, under the Rankine-Hugoniot jump conditions,
gas heats up from precursor temperatures of $T\sim10^4$ K to
anywhere from $10^5$ to $10^8$ K. Because the gas does not behave isothermally, even after
radiative cooling, shocked gas can maintain much higher equilibrium temperatures
in excess of several tens of kK, which can augment temperature diagnostic
features like the [\ion{O}{iii}] \W4363 auroral line
\citep[e.g.,][ Flury et al. in prep]{Riffel2021}, particularly if magnetic
fields or a paucity of metals inhibit cooling. In cases of a shock +
photoionization mixture, this temperature excess can lead to severe
over-estimations of the ionic emissivities due to their exponential dependence
on the measured electron temperature, which in turn leads to substantial
under-estimations of chemical abundances.
While some observations suggest a steep decline in electron temperature between
O$^{+2}$ and lower ionization species like O$^{+}$ and $\rm N^{+}$
\citep[e.g.,][Flury et al. in prep]{Riffel2021}, the temperature gradient in shocked gas
remains empirically unconstrained. As such, the temperature relations typically
invoked \citep[e.g.,][]{ArellanoCordova2020,Rogers2021} to infer electron
temperatures for ions without observed auroral lines could lead to inaccurate
ionic abundance estimates. While \no\ measured from $\rm N^{+}/O^+$ is
relatively insensitive to temperature, the use of \ion{N}{iv} with \ion{O}{iii}
to obtain the ionic abundances at $z>5$ can also suffer systematic biases from
shock-heating effects.

\subsection{Shock Compression}

Rankine-Hugoniot jump conditions predict compression of an adiabatic gas by a factor of 4 at the shock front, with an isothermal gas compression approaching a compression factor of $\infty$ \citep[e.g.,][]{source:osterbrock2006}. The rapid cooling processes which occur after the shock front (i.e., through the shock) cause the gas to compress by as much as 4 dex while the gas cools and enters the $10^4$ K recombination wake. At high velocities where the gas heating is much higher ($T_s\propto v_s^2$), rapid cooling causes the total shock to behave more isothermally, which can increase the gas compression rate by orders of magnitude \citep[e.g.,][]{Sutherland2017}. We find typical compression factors of 2 dex across all shock models with variations depending on magnetic field strength, shock velocity, and density \citep[cf.][\S8, Figures 8-9]{Sutherland2017}.
Magnetic fields can mitigate the compression effects of rapid cooling by exerting an outward magnetic pressure as the gas compresses, thereby reducing the radiative efficiency of the shock \citep[e.g.,][]{Draine1993}; however, even for strong magnetic fields ($B\ga10\rm\mu G$), we find that even low density gas ($1$ \ccm) with $v_s=100$ \kms\ can readily achieve a 2 dex increase in density.
Observations indicate that shock excitation is consistent with much higher electron densities \citep[e.g.,][]{Ho2014}. Given gas densities as high as $n_e\sim10^5$ \ccm\ or more observed in some cases at $z>5$
\citep[e.g.,][]{Maiolino2024,Topping2024b}, the post-shock densities of anywhere from $\sim10^5$ to $10^8$ \ccm\ due to compression of $\sim10^4$ \ccm\ gas are reasonable conditions to expect even at lower shock velocities of 100 \kms.

Among the forbidden and semi-forbidden transitions, the low $\rm ^4P$ states of the \ion{N}{iii}] quintuplet are uniquely insensitive to collisional de-excitations and may experience a corresponding flux enhancement through collisional excitation.
The comparably insensitive $\rm^5S_2$ state of the \ion{O}{iii}] lines has a higher critical density with two readily excited states immediately below related to the optical [\ion{O}{iii}] doublet and auroral lines, making the \ion{O}{iii}] doublet less augmented by density than the \ion{N}{iii}] quintuplet. Thus, the reported nitrogen ``excess'' may be due in part to gas density enhancement via compression by shocks. The apparent suppression of \ion{C}{iii}] emission in nitrogen-loud galaxies is likewise consistent with shock compression given the substantially low critical density of the $\rm^3P_2$ state. Properly accounting for these density effects may resolve the discrepancy between the C/N observed at $z\ga5$ with \emph{JWST} and expectations from standard nucleosynthesis models \citep[][ Arellano-Cordova et al. in prep a]{ArellanoCordova2022,Topping2024b, Curti2024, MarquesChaves2024}.

Dense gas conditions produced by shock compression, in particular lower velocity shocks, can provide the unique conditions for collisional suppression of \ion{C}{iii}] and excitation of the \ion{N}{iii}]. \ion{N}{iv} serves primarily as an ionization diagnostic as it is only slightly less sensitive to density than \ion{C}{iii}]. Gas in the early Universe can be substantially more dense \citep[e.g.,][]{Maiolino2024,Topping2024a}, which may be facilitated by shock compression of the gas. Both Mrk 996 and the Sunburst arc exhibit unusually high gas densities as well \citep[$\ga10^6$ \ccm\ and $\ga10^5$ \ccm, respectively,][]{Thuan2008,James2009,Pascale2023}, further indicating density enhancement that can be explained by shock compression.

\subsection{Reducing the Nitrogen Excess}

Radiative shocks can account for much of the observed strengths of UV nitrogen lines, reducing the reported excesses in nitrogen abundance to levels consistent with those observed in the local Universe. Such an adjustment would not eliminate the possibility of surplus nitrogen; however, the reduced excess removes the necessity of invoking more complex and/or fine-tuned scenarios to account for the measured abundances. High \no\ is reported for the Sunburst arc only when including the \ion{N}{iii}] quintuplet and is otherwise ``typical'' \citep{Pascale2023}, which may indicate density-augmented \ion{N}{iii}] flux. We suggest that shock compression could increase the gas density sufficiently to enhance \ion{N}{iii}], which in turn leads to over-estimates of \no\ in this object.
Indeed, \citet{Castellano2024} find that assuming high densities of $n_e\sim10^5$ \ccm\ and higher temperatures of $T_e\sim3\times10^{4}\rm~K$, conditions predicted for post-shock gas in many of our models, can reduce \no\ by at least $\sim0.1$ dex. In addition to high \no\ values, \citet{Topping2024b} find high densities ($n_e\sim10^5$ \ccm), high temperatures ($T_e\sim2.4\times10^4\rm~K$), and exceptionally weak [\ion{O}{ii}] \WW3726,29 lines, which could be explained by suppression of [\ion{O}{ii}] via collisional de-excitation of its upper $\rm^2D$ states (see Table \ref{tab:critDens}). However, we note that \citet{Castellano2024}, \citet{Topping2024b,Topping2024a}, and others compare $\rm N^{+2}+N^{+3}$ to $\rm O^{+2}$ to obtain \no\ and do not include an ionization correction factor to account for missing ion species $\rm N^{+}$, $\rm O^{+}$, and $\rm O^{+3}$. These corrections are necessary as indicated by the ionization potentials in Table \ref{tab:ionPot} and would further lower the apparent nitrogen excess by fully accounting for the oxygen abundance. Moreover, an appropriately calibrated temperature scaling relation could improve measurements of ionic abundances, thereby increasing the measured \oh\ and the expected \no\ values \citep[e.g.,][Flury et al. in prep]{Dors2020}

To test the degree to which the nitrogen excess persists in these objects, we
rerun a subset of shock models in the low-velocity, low metallicity regime with
a 0.5 dex increase in N/O with respect to the \citet{Nicholls2017} trend. We
find that we are able to reproduce the majority of the emission line fluxes of
nitrogen-excess galaxies with this 0.5 increase, which is consistent with
observations of stars and low-metallicity SFGs in the local Universe
\citep[e.g.,][Arellano-Cordova in prep.]{vanZee2006,Guseva2011,Stephenson2023}.
A 0.5 dex increase in N/O can arise from rapid mixing of Wolf Rayet ejecta with
the ISM following standard stellar nucleosynthesis abundance yields
\citep[e.g.,][]{Higgins2024}, an effect observed in nearby starburst galaxies
like NGC 5253
\citep[e.g.,][]{Kobulnicky1997,LopezSanchez2007,AbrilMelgarejo2024} and
ESO 338-IG04 \citep[e.g.,][]{Bik2018}. The nitrogen-enhanced knot of the
Sunburst arc has a stellar populations within the 3-6 Myr age range associated
with Wolf-Rayet stars
\citep{RiveraThorsen2017,RiveraThorsen2024,Vanzella2020,Pascale2023}
while Mrk 996 exhibits strong Wolf-Rayet features
\citep{Thuan1996,Thuan2008,James2009,Telles2014}, suggesting that Wolf-Rayet
stars are simultaneously enriching the ISM and driving winds which trigger the
radiative shocks. Local dwarf starburst galaxies like ESO 338-IG04 exhibit Wolf-Rayet stars causing simultaneous nitrogen enrichment and wind-driven shocks \citep{Bik2018}, which provides direct evidence for such scenarios in massive star clusters.

\section{Future Considerations for Models}

\subsection{Shocked Photoionized Gas}

Previous studies of shock-SFG ``composites'' have simply co-added the fluxes to illustrate a fractional contribution of each component to the total H$\beta$ flux \citep[e.g.,][]{Jaskot2016,Chisholm2024}.
While such an approach is physical when shocks and photoionized regions
simply fall in the same spectroscopic aperture, the sensitivity of radiative
shocks to the precursor conditions precludes such an approach when the gas
is simultaneously illuminated by a photoionizing SED independent of the shock. In the context of the
bright nitrogen lines, a scenario where, e.g.,  Wolf Rayet stars are photoionizing
gas which is subsequently shocked by a stellar wind or an AGN is photoionizing gas which is subsequently shocked by a radiatively driven wind \citep[e.g.,][]{Riffel2021}, a shock + photoionization
mixture must be treated simultaneously within a single model \citep[e.g.,][]{Dors2021}. Such an approach
may explain the particularly high \ion{N}{iii}]/\ion{C}{iii}] ratios observed
for AGN candidates GNz11 and GS 3073 and could account simultaneously for heating
and ionization of the precursor by the AGN and gas heating, ionization, and
compression by the shock. Future work should entail detailed treatment of these
combined effects in order to reproduce the extreme $z>5$ galaxies, particularly those with AGN signatures.

\subsection{Dust}

The dust content of galaxies at $z>5$ is disputed. Signatures of dust-poor or
even dust-free galaxies are present in the early Universe
\citep[e.g.,][]{Cullen2024,Saxena2024}; however, evidence for dust in SFGs
persists out to at least $z\sim8$--even as early as $z\sim12$ or more--with
relatively weak evolution with redshift \citep[e.g.,][]{Ciesla2024,Saxena2024}. While our
primary shock models contain no dust, our additional dust-rich models suggest its effects may be important to consider for
a variety of physical reasons.


Full destruction of the dust grains by the hot post-shock conditions may not be
a reasonable assumption as anywhere from 10 to 70\% grains can survive the shock
\citep[e.g.,][]{Borkowski1995,Jones1996,Dopita2016}. In cases of low-velocity
shocks, models indicate that the rate of grain destruction decreases due to the
decline in thermal sputtering and shattering \citep[e.g.,][]{McKee1987}. Dust
grains can dramatically impact shocks since they are coupled to gas via drag and
to the magnetic field via charge and may also serve as an important source of
radiative cooling \citep{Draine1993}. Therefore, the incorporation of dust,
including the various destruction mechanisms, into shock models should be
addressed in greater detail.

\subsection{Catastrophic Cooling}

Two objects we have considered in this study exhibit evidence for catastrophic cooling in addition to possible shock excitation: the Sunburst arc \citep{Pascale2023} and Mrk 71 \citep{Oey2023,Smith2023}. Catastrophic cooling can occur in shocks \citep[e.g.,][]{Falle1975,Falle1981,Smith1989} and can be associated with phenomena such as supernova remnants \citep[e.g.,][]{Falle1975,Falle1981,Smith1989,Silich2007} and winds \citep{Smith1989,Wang1995,Silich2004,Silich2007,Gray2019}. Catastrophic cooling is more likely to occur in dense gas environments due to over-pressurization \citep[e.g.,][]{Silich2004}, which may be facilitated by shocks and made more common in denser gas environments such as those in Green Pea galaxies \citep[e.g.,][]{Jaskot2019} or those at high redshift \citep[e.g.,][]{Jones2024,Maiolino2024,Topping2024a,Zavala2024a,Zavala2024b}. Shocks could further contribute to catastrophic cooling via Vishniac instabilities wherein small perturbations cause fragmentation of the shock into dense cloudlets \citep{Vishniac1983}. Therefore, the effects of catastrophic cooling may be important for interpreting observations of galaxies with shock-like line emission.

We have compared archival catastrophic cooling predictions from \citet{Danehkar2022} to our models and find that they are similarly capable reproducing the augmented \ion{N}{iii}] \WW1747-54 quintuplet but do not predict similar suppression of the [\ion{C}{iii}] \W1907 line. We note that catastrophic cooling libraries are, to date, quite novel and as a result might not span the same range in ionization parameter, ionizing source, or metallicity as our or other models. Indeed, the available models, while immensely useful, apply strictly to catastrophic cooling in galactic winds. Future work should include more detailed, robust, and extensive catastrophic cooling model line predictions in a wider range of scenarios. If catastrophic cooling in shocks can explain the observations of galaxies at $z>5$, it may also indicate positive feedback in SFGs by suppressing winds and forming dense clouds \citep[e.g.,][]{Silich2007,Jaskot2017} and could even promote Lyman continuum escape \citep[e.g.,][]{Jaskot2019,Flury2024}.


\section{Conclusion}\label{sec:conclude}

To assess the origins of the putative nitrogen excess observed in galaxies at $z>5$, we compute a series of new excitation models for SFGs, AGN, and shocks using {\tt MAPPINGS} with empirical chemical abundance and dust depletion patterns and new atomic data shared across all models to mitigate any systematic differences. We define excitation regimes by exploring a wide range of ionizing SEDs from stellar populations and AGN as well as a variety of shock conditions.

We assemble flux ratio diagrams motivated by a combination of observational and physical constraints:
\begin{itemize}
    \item detectability of emission lines
    \item resolvability of emission lines
    \item insensitivity to extinction
    \item sensitivity to SED shape
    \item sensitivity to gas density
    \item insensitivity to metallicity
\end{itemize}
We compare emission line flux ratios from spectra of nitrogen-excess galaxies observed by \emph{JWST} with our excitation diagnostics. From our assessment, we conclude that galaxies exhibiting unusually bright \ion{N}{iii}] and/or \ion{N}{iv}] have spectra consistent with low-velocity shocks and cannot be explained by other phenomena such as photoionization by AGN or stellar populations. These nitrogen-excess galaxies, along with several reference galaxies, consistently fall in diagnostic regions associated with shock ionizaiton when considering \ion{C}{iii}]/\ion{He}{ii}, \ion{C}{iv}/\ion{He}{ii}, \ion{C}{iv}/\ion{C}{iii}], and \ion{O}{iii}]/\ion{He}{ii}.

Of the diagnostics considered here, the \ion{C}{iii}] / \ion{He}{ii} vs \ion{O}{iii}] / \ion{He}{ii} and \ion{N}{iii}] / \ion{C}{iii}] vs \ion{O}{iii}] / \ion{He}{ii} diagrams best distinguish between shock ionization and photoionization, predominantly owing to ionization and compression effects from the shock. To distinguish between different photoionization mechanisms, we recommend \ion{N}{iv}] / \ion{C}{iii} vs \ion{O}{iii}] / \ion{He}{ii} when \ion{O}{iii}] can be resolved from \ion{He}{ii}. If \ion{He}{ii} is not resolved from \ion{O}{iii}], the \ion{C}{iii}] / \ion{C}{iv} vs \ion{C}{iv} / (\ion{He}{ii}+\ion{O}{iii}) can be used to distinguish SFGs from shocks and most AGN scenarios with the caveat that the \ion{C}{iv} doublet can suffer from stellar contamination.

We find an exceptional ability of \ion{N}{iii}] and \ion{N}{iv}] to distinguish between shock and AGN excitation, defining a unique diagnostic space for shocks. Due to the relatively constant C/N with respect to \oh\ and metallicity in general, the \ion{N}{iii}]/\ion{C}{iii}] and \ion{N}{iv}]/\ion{C}{iii}] diagnostic flux ratios provide an optimal means for distinguishing ionizing mechanisms.
Both the nitrogen-excess galaxies and the reference galaxies consistent with shocks fall in the model-defined region for shocks, confirming the success of \ion{N}{iii}] and \ion{N}{iv}] as shock diagnostics. To demonstrate the shock nature of these galaxies, we find kinematic evidence in the reference galaxies indicative of winds with velocities of 100 - 200 \kms. These velocities are consistent with the shock velocities of models which best describe the emission line flux ratios, corroborating our interpretation that shocks are a key ionizing mechanism in these objects.

The clear evidence for shock excitation suggests complications in the measurement of ``direct method'' nitrogen and oxygen abundances due to compression, heating, and ionizaiton of gas by shocks. Moreover, the nitrogen abundances of these galaxies need not be extreme in order to agree with ionization models, with \no\ 0.2-0.3 dex higher than expected at the reported \oh. In even the most extreme cases, such as CEERS 3073 and GN-z11, the nitrogen enhancement may well be only 0.5 dex, still within the dispersion of local extragalactic observations \citep[e.g.,][ Arellano-Cordova et al. in prep.]{vanZee2006,Guseva2011,Berg2012}. We interpret the implied 0.3 excess in nitrogen, combined with low velocity shocks, as evidence for enrichment and excitation via winds from Wolf Rayet stars. Proper consideration of the shock ionization and its effects on the measured emission line fluxes can reconcile observations with conventional chemical evolution models and abundance patterns measured in the local Universe.

\section*{Acknowledgements}

We thank the anonymous referee for their insightful comments.

SRF acknowledges support from NASA/FINESST 80NSSC23K1433. SRF thanks C Morisset and A Alarie for insight regarding the implementation of {\tt MAPPINGS V}, M Mingozzi for assistance with the CLASSY observations, and K Duncan for suggestions regarding optimal release of the models.

This research is based in part on observations made with the NASA/ESA Hubble Space Telescope obtained from the Space Telescope Science Institute, which is operated by the Association of Universities for Research in Astronomy, Inc., under NASA contract NAS 5–26555. These observations are associated with programs GO 15840, 16213, 16261.

This work is based in part on observations made with the NASA/ESA/CSA James Webb Space Telescope. The data were obtained from the Mikulski Archive for Space Telescopes at the Space Telescope Science Institute, which is operated by the Association of Universities for Research in Astronomy, Inc., under NASA contract NAS 5-03127 for JWST. These observations are associated with programs GTO 1180, 1181, 1210, 1286, GO 1895, 1963, 3215. The authors acknowledge the JADES and CEERS teams for developing their observing programs with a zero-exclusive-access period.

This publication made use of the {\tt MAPPINGS V} software, which utilizes the {\tt CHIANTI} database. {\tt CHIANTI} is a collaborative project involving George Mason University, the University of Michigan (USA), University of Cambridge (UK) and NASA Goddard Space Flight Center (USA).

We acknowledge the use of the following software, repositories, and other digital resources:
{\tt BPASS} \citep{bpass2.2},
{\tt CHIANTI} \citep{chianti10},
{\tt EOMS} \citep{Alarie2019},
{\tt MAPPINGS V} \citep{Sutherland2017,mappingsV},
{\tt matplotlib} \citep{matplotlib},
{\tt numpy} \citep{numpy},
{\tt oxaf} \citep{optxagnf},
{\tt scipy} \citep{scipy},
{\tt Starburst99} \citep{starburst99v1,starburst99v2,starburst99v2a},
{\tt vygrboi} (\url{github.com/sflury/vygrboi})

For the purpose of open access, the author has applied a Creative Commons Attribution (CC BY) licence to any Author Accepted Manuscript version arising from this submission.

\section*{Data Availability}

The predicted emission line fluxes, abundance patterns, gas conditions, and properties of key ions for the models presented here will be made available online at \url{http://dx.doi.org/10.5281/zenodo.14140949}. Ionizing SEDs not publicly attainable elsewhere will be made available on request.




\bibliographystyle{mnras}
\bibliography{bib,soft} 






\bsp	
\label{lastpage}
\end{document}